\documentclass[twocolumn,amsmath,amssymb,aps, prb, superscriptaddress,
showkeys,showpacs,nofootinbib] {revtex4-1}
\usepackage{graphicx}
\usepackage{bm}
\usepackage[colorlinks=true,allcolors=blue]{hyperref}
\usepackage{dcolumn}

\usepackage{float}
\allowdisplaybreaks
\begin{document}
\title{Quantum phases of a one-dimensional Majorana-Bose-Hubbard model}
\author{Ananda Roy}
\email{ananda.roy@tum.de}
\affiliation{Department of Physics, T42, Technische Universit\"at Mu\"nchen, 85748 Garching, Germany}
\author{Johannes Hauschild}
\affiliation{Department of Physics, T42, Technische Universit\"at Mu\"nchen, 85748 Garching, Germany}
\author{Frank Pollmann}
\affiliation{Department of Physics, T42, Technische Universit\"at Mu\"nchen, 85748 Garching, Germany}
\affiliation{Munich Center for Quantum Science and Technology (MCQST), 80799 Munich, Germany}
\begin{abstract}
Majorana zero modes (MZM-s) occurring at the edges of a 1D, p-wave, spinless superconductor, in absence of fluctuations of the phase of the superconducting order parameter, are quintessential examples of topologically-protected zero-energy modes occurring at the edges of 1D symmetry-protected topological phases. In this work, we numerically investigate the fate of the topological phase in the presence of phase-fluctuations using the density matrix renormalization group (DMRG) technique. To that end, we consider a one-dimensional array of MZM-s on mesoscopic superconducting islands at zero temperature. Cooper-pair and MZM-assisted single-electron tunneling, together with finite charging energy of the mesoscopic islands, give rise to a rich phase-diagram of this model. We show that the system can be in either a Mott-insulating phase, a Luttinger liquid (LL) phase of Cooper-pairs or a second gapless phase. In contrast to the LL of Cooper-pairs, this second phase is characterized by nonlocal string correlation functions which decay algebraically due to gapless charge-$e$ excitations. The three phases are separated from each other by phase-transitions of either Kosterlitz-Thouless or Ising type. Using a Jordan-Wigner transformation, we map the system to a generalized Bose-Hubbard model with two types of hopping and use DMRG to analyze the different phases and the phase-transitions.%
\end{abstract}
\maketitle 

\section{Introduction}
Topological phases of matter cannot be characterized by a local order parameter and fall outside the celebrated symmetry-breaking paradigm of Landau. Investigation of these novel phases of matter has been central to a better comprehension of collective behaviors of strongly interacting many-particle systems. Certain topological phases in 2D have been predicted to host quasiparticles with anyonic exchange statistics, which are valuable for quantum computation.~\cite{Kitaev2003, Kitaev2006, Levin2005, Levin2006, Nayak2008} For the latter, Majorana zero modes (MZM-s) occurring in vortex defects in 2D chiral p-wave superconductors~\cite{Read2000, Kitaev2006} are promising candidates.  Braiding of these MZM-s, together with `magic state distillation' can be used to perform the operations necessary for universal quantum computation.~\cite{Bravyi2006} These MZM-s are also predicted to occur at the edges of a 1D spinless, p-wave superconductor.~\cite{Kitaev2001} In the latter incarnation, a 2D array of these MZM-s on mesoscopic superconducting islands gives rise to the $\mathbb{Z}_2$ toric code~\cite{Xu2010, Terhal2012, Landau2016, Karzig2016, Roy2017}, one of the most promising platforms for quantum computation. 

MZM-s are the key feature of the topological phase of a 1D spinless, p-wave superconductor. However, in modeling the latter, it was assumed that the 1D superconductor is in contact with a bulk 3D superconductor which removes the phase-fluctuations of the superconducting order parameter and suppresses phase-slip rates exponentially with the system size\cite{Fidkowski2011}. This raises the question: does the topological phase survive when these assumptions are relaxed? This question has previously been addressed in the context of the existence of topological-protection of MZM-s in a number-conserving theory using field theory (bosonization) computations~\cite{Fidkowski2011, Sau2011} (see also Refs. \onlinecite	{Cheng2015, Knapp2018, Knapp2019}). Furthermore, number-conserving Hamiltonians giving rise to topologically-protected edge-modes have also been proposed.~\cite{Ortiz2014, Iemini2015} 

In this work, we numerically investigate the fate of the topological phase when the 1D spinless, p-wave superconductor is in contact with a 1D s-wave superconductor using the density matrix renormalization group (DMRG) technique. We model the 1D s-wave superconductor by a chain of mesoscopic superconducting islands separated by tunnel junctions. Each mesoscopic island contains two MZM-s. 
Cooper pairs tunnel between the superconducting islands at a rate $E_J$, while the presence of MZM-s leads to the coherent tunneling of single electrons at a rate $E_M$. Furthermore, the mesoscopic nature of each island gives rise to a finite charging energy, set by the scale: $E_C = (2e)^2/2C$ where $C$ is the self-capacitance of each island to a common ground. It is this charging energy of the islands that gives rise to fluctuations of the phase of the superconducting order parameter. Throughout, we assume a homogeneous array without any disorder and neglect inter-island capacitances. As we will show in this work, the phase-diagram of this model -- hereafter referred to as the Majorana-Bose-Hubbard (MBH) model --  is rich and various limiting cases of this model have been analyzed before. This is explained below.  

First, consider the limit of vanishing single electron tunneling ($E_M=0$). In this case, the MZM-s decouple completely from the remaining degrees of freedom and play no role in the phase-diagram. The resulting model is the Bose-Hubbard model in the limit of high-occupancy of each site.~\cite{Fisher1989, Fazio2001} The Cooper-pairs tunneling between the islands play the role of the bosons hopping between the sites, while the charging energy of each island is the onsite-repulsion. The role of the chemical potential is played by a gate-voltage on each island. The phase-diagram of this model is well-established.~\cite{Bradley1984, Korshunov1989, Glazman1997} For $E_J\gg E_C$, the system is a 1D superconductor or a charge-$2e$ Luttinger liquid (LL). The latter phase is characterized by the power-law correlation of the Cooper-pair creation and annihilation operators, the exponent of the power law depending on the Luttinger parameter. Lowering $E_J/E_C$ causes the system to undergo a quantum phase-transition of Kosterlitz-Thouless (KT) type into a Mott-insulating (MI) phase. In the MI phase, the charge on each island is pinned to an {\it integer} value in units of Cooper pairs. Physically, this transition is driven by an unbinding of $2\pi$ quantum phase-slips. The latter causes tunneling between the minima of the Josephson potential and can be viewed as vortices in space-time. These vortices become free in the MI phase.~\cite{Fazio2001} Equivalently, one can map the problem to a 2D classical XY model at finite temperature using the transfer-matrix method.~\cite{Kogut1979} In this classical picture, the LL phase corresponds to the (low temperature) superconducting phase. The transition is then driven by the unbinding of $2\pi$ vortices into a (high temperature) resistive phase.~\cite{Bradley1984}

Now consider the limit of vanishing Cooper-pair tunneling ($E_J = 0$). In this case, there is no tunneling of Cooper pairs and only single electrons tunnel between neighboring islands. For $E_M\gg E_C$, the system is in a gapless phase.  However, in contrast to the charge-$2e$ LL of Cooper-pairs, this phase is characterized by nonlocal string correlation functions~\cite{Pollmann2012, Bahri2014}, which decay algebraically in this phase.~\cite{Xu2010} As will be explained later, this algebraic decay is due to gapless charge-$e$ excitations and is generic for phases in the proximity of a topological phase. Lowering $E_M/E_C$ causes the system to undergo a quantum phase-transition to a MI phase. In the MI phase, the charge on each island is pinned to {\it half-integer} values in units of Cooper pairs. This is because unpaired electrons can be hosted in the fermionic modes built out of the MZM-s. The nature of the phase-transition is again KT, but this transition is driven by an unbinding of $4\pi$ quantum phase-slips. This is best seen using a Jordan-Wigner (JW) transformation which reduces this model to a variation of the Bose-Hubbard model.~\cite{Xu2010, Roy2018a} In this model,  charge-$e$ bosons hop between the mesoscopic islands. The nonlocal string correlation function is then mapped to a local correlation function of charge-$e$ bosons and in this JW-transformed picture, the gapless phase can be thought of as a charge-$e$ LL. Thus, we will use this name to refer to it. For this limit, while some aspects of the phase-diagram have been predicted,~\cite{Xu2010} several questions regarding characterization of the two phases remain open. 
 
The main goal of this work is to perform a numerical analysis of the different phases and the phase-transitions of the MBH model using DMRG. We consider the general  case of both $E_J, E_C$ being nonzero. Using a JW transform, we map the model to a generalized Bose-Hubbard model with hopping of bosons carrying either charge $e$ or $2e$, on which DMRG is performed. From the DMRG analysis of the JW-transformed model, we straightforwardly infer the properties of the fermionic model. As will be shown in this work, this model can be mapped to a generalized classical XY model with $2\pi$ and $4\pi$ vortices~\cite{Lee1985, Serna2017} using the transfer-matrix method. This will allow us to verify several of the DMRG results concerning the phase-diagram by comparing to the Monte Carlo results of the classical model~\cite{Serna2017}. While the phase-transitions remain the same under this transfer-matrix mapping, the properties of the different phases of the MBH model are naturally completely different from the classical model. We show that depending on the ratios $E_J/E_C$ and $E_M/E_C$, the system can be either a MI, a  charge-$2e$ LL or a charge-$e$ LL. The three phases can be distinguished as follows (see  Table \ref{tab1} for more details). Generalizing the nonlocal string correlation functions  proposed in Ref.~\onlinecite{Bahri2014} to include phase-fluctuations, we show that the charge-$e$ LL phase is characterized by an algebraic decay of a nonlocal correlation function. The latter comprises a string of bosonic operators terminated by fermionic ones. In contrast, this nonlocal correlation function decays exponentially in the MI and the charge-$2e$ LL phases. The algebraic vs. exponential decay directly follows from the gapless vs. gapped nature of the charge-$e$ excitation spectrum. The MI and the charge-$2e$ LL phases are then distinguished by the Cooper-pair correlation function, which decays exponentially in the MI phase and algebraically in the charge-$2e$ LL phase, the last algebraic decay being a signature of the gapless $2e$ excitation spectrum. Note that the Cooper-pair correlation function also decays algebraically  in the charge-$e$ LL phase. This is because the gapless charge-$e$ excitation spectrum necessarily implies a gapless $2e$ excitation spectrum. 

After characterizing the different phases, we analyze the phase-transitions that separate them. As expected, we find the transition between the charge-$2e$ LL and the MI phases to be of KT type. The transition from the two LL phases is of Ising type. As the system transitions between the two phases both with gapless $U(1)$ degrees of freedom, an emergent $\mathbb{Z}_2$ degree of freedom turns gapless, causing the transition.  Physically, this Ising degree of freedom transforms the $2\pi$-periodic charge-$2e$ LL phase to a $4\pi$-periodic charge-$e$ LL phase.  We note that similar results have been previously obtained for a 2D version of this model~\cite{Roy2017}, a generalized classical XY  model~\cite{Serna2017} as well as for several lattice gauge theory models~\cite{Fradkin1979, Senthil2000}. Interestingly, the nature of the transition from the charge-$e$ LL to the MI phase depends on the magnitude of $E_J/E_C$. The Ising transition line that separates the two LL phases continues to separate part of the MI phase from the charge-$e$ LL phase. At this somewhat remarkable phase-transition, occurring at $E_J/E_C\sim1$, both the  U(1) and $\mathbb{Z}_2$ degrees of freedom simultaneously undergo long-wavelength fluctuations. This result is consistent with the Monte Carlo simulations of the classical generalized XY model~\cite{Serna2017}.  As $E_J/E_C$ is lowered to 0, this Ising transition line turns into a KT transition line. Then the latter  separates the charge-$e$ LL and the MI phases. It is this KT type transition that can be understood as the usual unbinding of $4\pi$ quantum phase-slip transition and was predicted in Ref.~\onlinecite{Xu2010}. 

The paper is organized as follows. In Sec. \ref{model}, we introduce the MBH model. Sec. \ref{fermionic_model} discusses the fermionic version of the model and its phases, while Sec. \ref{bosonic_model} describes model after JW transformation. In Sec. \ref{phasediag}, we provide an intuitive picture of the expected phase-diagram. Sec. \ref{dmrg} provides our DMRG results. In Sec. \ref{topornot}, we discuss to what extent the charge-$e$ LL phase can be considered as a topological phase. In Sec. \ref{concl}, we summarize our results and provide concluding perspectives. Appendix \ref{2maj} discusses the generalization of our model to four MZM-s on each superconducting island. Appendix \ref{extra_DMRG} provides additional details concerning the DMRG simulations.

\section{The Majorana-Bose-Hubbard Model}
\label{model}
\subsection{Description of the model}
\label{fermionic_model}
\begin{figure}
\includegraphics[width = 0.48\textwidth]{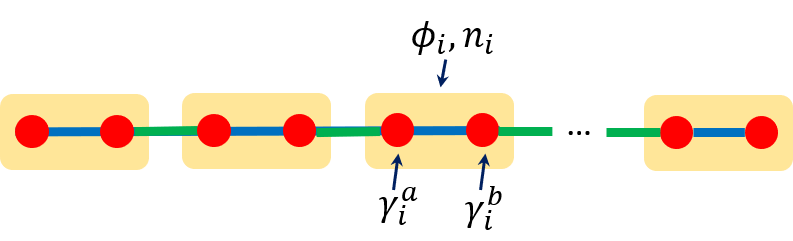}
\caption{\label{1darray} Schematic of a 1D array of mesoscopic superconducting islands (denoted by yellow boxes) with MZM-s (denoted by red dots). In addition to the rotor degrees of freedom, $\phi_i, n_i$, each island has two MZM-s, denoted by $\gamma_i^a, \gamma_i^b$. }
\end{figure}

Consider a 1D array of $L$ mesoscopic superconducting islands, each having the same charging energy, $E_C$ (see Fig.~\ref{1darray}). The superconducting phase and the excess number of fermions, in units of Cooper pairs, on each island are denoted by $\phi_i$ and $n_i$, which obey canonical conjugate commutation relations: $[\phi_i, n_j] = i\delta_{i,j}$, $i,j = 1, \ldots, L$. The exponential of the phase-operators are bosonic operators which alter number of the fermions on each island by one Cooper-pair: $e^{\pm i\phi_i}|n_i\rangle = |n_i\pm1\rangle$. In what follows, we will denote them  as $b_{i}  = e^{-i\phi_i}$, $b_{i}^\dagger  = e^{i\phi_i}$. On each island, there are two MZM-s, denoted by $\gamma_i^a, \gamma_i^b$, which obey the fermionic commutation relations: $\{\gamma_i^\alpha,\gamma_j^\beta\} = 2\delta_{i,j}\delta_{\alpha,\beta}$, $i,j = 1,\ldots, L$ and $\alpha, \beta = a,b$. 

The Hamiltonian describing the system is given by 
\begin{align}
\label{ham2maj}
H &= H_C + H_J + H_M + H_g,\\\label{ham2maj_a}
H_C &= E_C\sum_{i=1}^L n_i^2, \\\label{ham2maj_b} H_J &= -E_J \sum_{i=1}^{L-1}\cos(\phi_i - \phi_{i+1}),\\\label{ham2maj_c}
H_M &= E_M\sum_{i=1}^{L-1}i\gamma_i^b\gamma_{i+1}^a\cos\frac{\phi_i-\phi_{i+1}}{2},\\\label{ham2maj_d} H_g &= -E_g\sum_{i=1}^L n_i.
\end{align}
Here, $E_J$ ($E_M$) denote the rate of tunneling of Cooper-pairs (single-electrons) and $E_g$ is the gate voltage. Clearly, the Josephson tunneling term [Eq.~\eqref{ham2maj_b}] denotes a boson-hopping term $b_{i}b_{i+1}^\dagger + {\rm{H.c}}$. The single-electron tunneling term [Eq.~\eqref{ham2maj_c}] is more complex. The factor of 1/2 in the cosine argument indicates that one-half of the charge of the Cooper-pair hops between neighboring islands, while the fermionic operators keep track of the change of the fermion number parity associated with the transfer of these electrons.  The fermion parity on each island is given by $P_i = i\gamma_i^a\gamma_i^b$. The physical Hilbert space is spanned by wavefunctions satisfying~\cite{Fu2010}
\begin{equation}
\label{parconstr}
\psi(\phi_i + 2\pi) = e^{2\pi in_i}\psi(\phi_i) = P_i\psi(\phi_i).
\end{equation}
Note that the wave-function and the Hamiltonian are invariant under $\phi\rightarrow \phi + 4\pi$. This implies that the conjugate variables, $n_i$-s, can be half-integers. Physically, $n_i$ is a half-integer when there is an unpaired electron on the $i^{\rm{th}}$ island occupying the fermionic mode built out of the two MZM-s $\gamma_i^a, \gamma_i^b$. 
As explained in the introduction, for $E_J, E_M\ll E_C$, the system is a MI. Increasing $E_J, E_M$ causes the system to undergo a quantum phase transition to either a charge-$2e$ LL or a charge-$e$ LL. To distinguish the different phases, we introduce the following correlation functions. First, consider the correlation function of Cooper-pair creation and annihilation operators:
\begin{equation}
\label{sb}
S_{b}(i,j) = \big\langle b_i b_j^\dagger\big\rangle.
\end{equation}
In the MI phase, the charge-$2e$ excitation spectrum is gapped. Hence, $S_b$ decays exponentially with the separation $|i-j|$. However, the charge-$2e$ excitation spectrum is gapless in both the charge-$2e$ LL and the charge-$e$ LL. As a result, $S_b$ decays algebraically in these phases. Next, we define two further nonolocal string correlation functions: 
\begin{align}
\label{Striv}
S_{\rm{triv}}(i,j) &= \Bigg\langle \prod_{k = i+1}^{j} e^{2\pi i n_k}\Bigg\rangle = \Bigg\langle \prod_{k = i+1}^{j} P_k\Bigg\rangle,\\
\label{Stop}
S_{\rm{top}}(i,j) &= \Bigg\langle e^{-i\frac{\phi_i}{2}}i\gamma_i^b \Big(\prod_{k = i+1}^{j-1}e^{2\pi i n_k}\Big)e^{i\frac{\phi_j}{2}}\gamma_j^a\Bigg\rangle,
\end{align}
where the second equality of Eq.~\eqref{Striv} follows from Eq.~\eqref{parconstr}. These two nonlocal correlation functions are appropriate generalizations of the string correlation functions introduced in Ref. \onlinecite{Bahri2014} to account for phase-fluctuations~(hence, the inclusion of the appropriate phase-factors)~\footnote{These operators are related to the Fredenhagen-Marcu operators of coventional lattice gauge theory~\cite{Fredenhagen1983, Gregor2011, Ziesen2019}}. The correlation functions, $S_{\rm{triv}}$ and $S_{\rm{top}}$, are able to distinguish between topological and trivial phases. The value of $S_{\rm{triv}}$ ($S_{\rm{top}}$) is nonzero (zero) in a trivial phase and zero (nonzero) in a topological phase. We will use these order parameters to distinguish between the different phases of our model. However, in contrast to the case considered in Ref.~\onlinecite{Bahri2014}, it is not the asymptotic value of the correlation functions, but the nature of the decay that distinguishes the different phases of our model. Next, we summarize the behaviors of the two nonlocal correlation functions in the three phases. The detailed justification is given in the next subsection. 

The trivial correlation function, $S_{\rm{triv}}$, is nonzero in both the MI and the charge-$2e$ LL phases. On the other hand, $S_{\rm{triv}}$ decays algebraically to zero in the charge-$e$ LL phase. While at first sight, this might indicate that the charge-$e$ LL phase is topological, $S_{\rm{top}}$ indicates a more subtle behavior. For large enough distances, $S_{\rm{top}}$ decays to zero in {\it all} the three phases, but the nature of the decay is different. In the MI and the charge-$2e$ LL phase, $S_{\rm{top}}$ decays to zero exponentially, while in the charge-$e$ LL phase, the decay is algebraic. 
The two different types of decay are due to the gapped vs. gapless nature of the charge-$e$ spectrum. The characteristics of the different order parameters in the different phases are summarized in Table \ref{tab1}. Further discussion on the ramification of the algebraic nature of the decays of $S_{\rm{top}}$ and $S_{\rm triv}$ in the charge-$e$ LL phase is given in Sec.~\ref{topornot}. 

\begin{table}
\begin{ruledtabular}
\begin{tabular}{cccc}
phases&MI&charge-$2e$ LL&charge-$e$ LL\\\hline
$S_b$&exponential&algebraic&algebraic\\
$S_{\rm{triv}}$&does not decay&does not decay&algebraic\\
$S_{\rm{top}}$&exponential&exponential&algebraic
\end{tabular}
\caption{\label{tab1} Nature of the decay of the various order parameters in the three phases. The Cooper-pair correlation function, $S_b$ [Eq.~\eqref{sb}] decays exponentially (algebraically) in the MI phase (both the LL phases) since the $2e$ spectrum is gapped (gapless). The trivial correlation function, $S_{\rm{triv}}$ [Eq.~\eqref{Striv}], is nonzero and constant in both the MI and charge-$2e$ LL phases, while decaying algebraically in the charge-$e$ LL phase. Finally, the topological correlation function, $S_{\rm{top}}$ [Eq.~\eqref{Stop}], decays exponentially in both the MI and the charge-$2e$ LL phases, where the charge-$e$ spectrum remains gapped. The latter spectrum is gapless in the charge-$e$ LL phase, where $S_{\rm{top}}$ decays algebraically.}
\end{ruledtabular} 
\end{table}

\subsection{Mapping to a generalized Bose-Hubbard model with two types of hopping}
\label{bosonic_model}
To provide quantitative predictions of the characteristics of the phases and the phase-transitions that separate the different phases, it is convenient to use the JW transformed model. This is described next. As we will see, the nonlocal correlation functions have rather simple interpretations in the JW transformed model and provide justifications for the different behaviors of the various correlation functions given in the previous subsection.

We map the fermions to spins as follows: 
\begin{equation}
\label{jwtrans}
\gamma_i^a = \Big(\prod_{j<i}Z_j\Big)X_i, \ \gamma_i^b = \Big(\prod_{j<i}Z_j\Big)Y_i,
\end{equation}
where $X_i,Y_i$ and $Z_i$ are the Pauli operators, $i = 1,\ldots, L$. 
An additional unitary transformation leads to the wavefunctions to be $2\pi$ periodic:~\cite{VanHeck2012}
\begin{align}
\label{unitary}
H\rightarrow \Omega^\dagger H \Omega,\ \psi\rightarrow \Omega^\dagger\psi,\ \Omega = \prod_{i=1}^L e^{i(1-{P}_i)\phi_i/4}.
\end{align}
These manipulations transform the Majorana-assisted single-electron tunneling term of Eq.~\eqref{ham2maj_c} to a hopping of charge-$e$ bosons: $\tilde{b}_{i}\tilde{b}_{i+1}^\dagger + {\rm{H.c}}$, where $\tilde{b}_i = {b}_i\sigma_i^+ + \sigma_i^-$. \footnote{This transformation can be viewed as a different way to count charges on each island. Prior to the transformation, the spins carried no charge, but merely changed the fermion number parity. After this transformation, the charges are distributed across the operators $b_i, b_i^\dagger$ (create and destroy charge $2e$) and $\sigma_i^\pm$ (create and destroy charge $e$).} As a result, the Hamiltonian becomes that of a generalized Bose-Hubbard model with two types of hopping terms:
\begin{align}
\label{ham2maj_2a}
H_C &= E_C\sum_{i=1}^L \tilde{n}_i^2,\ H_J = -\frac{E_J}{2} \sum_{i=1}^{L-1}(b_{i}b_{i+1}^\dagger + {\rm{H.c}}),\\\label{ham2maj_2b}
H_M &= -\frac{E_M}{2}\sum_{i=1}^{L-1}(\tilde{b}_{i}\tilde{b}_{i+1}^\dagger + {\rm{H.c}}),\  H_g = -E_g\sum_{i=1}^L \tilde{n}_i.
\end{align}
In the above equation, $\tilde{n}_i$ denotes the excess number of fermions on the $i^{\rm{th}}$ island, which can be half-integers. It is given by $\tilde{n}_i = n_i + (1+Z_i)/4$, where $n_i$ is the excess number of Cooper pairs, the latter being integers. If there is an excess electron on the $i^{\rm{th}}$ island, then $Z_i = +1$ and $\tilde{n}_i$ is a half-integer, as expected. The following commutation relations hold between the operators $b_i, \tilde{b}_i$ and the number operators $\tilde{n}_j$:
\begin{equation}
[\tilde{n}_i, \tilde{b}_j^\dagger] = \frac{1}{2}\delta_{ij}\tilde{b}_j^\dagger,\ [\tilde{n}_i, b_j^\dagger] = \delta_{ij}b_j^\dagger.
\end{equation}
The transformations also make the symmetries of the model explicit. In addition to time-reversal symmetry, the model has a conserved $U(1)$ charge $\tilde{n}_i$. As explained above, physically, this charge is the electronic charge in units of Cooper-pairs. Note that there is also conservation of the overall parity: $\prod_{i=1}^L Z_i$. However, the conservation of the latter is {\it not} independent of the conservation of the $U(1)$ charge. 

Under the JW mapping, the order parameters introduced above are transformed as follows. The Cooper-pair correlation function, $S_b$, is left unchanged. On the other hand, $S_{\rm{triv}}$ and $S_{\rm{top}}$ are transformed to
\begin{align}
\label{Sts}
S_{\rm{triv}}(i,j) &= \Bigg\langle \prod_{k = i+1}^{j} Z_k\Bigg\rangle,\ S_{\rm{top}}(i,j) =\big\langle \tilde{b}_i \tilde{b}_j^\dagger\big\rangle,
\end{align}
where we have dropped overall minus signs which are not relevant for the analysis of the phase-diagram. From the above formulas, we can infer the behavior of the $S_{\rm{triv}}$ and $S_{{\rm top}}$. First, the trivial correlation function, $S_{\rm{triv}}$, since it is given by the exponential of the excess number of fermions ($\tilde{n}_i$) on each island, is pinned to a constant nonzero value in the MI phase. Second, in the charge-$2e$ LL phase, the excess number of fermions modulo 2 is still pinned since the system is a superconductor of Cooper pairs. This is because the charge-$e$ spectrum remains gapped. Third, $S_{\rm{triv}}$ decays algebraically when the charge-$e$ spectrum is gapless. A quantitative justification for this behavior is obtained by noting that $S_{\rm{triv}}$ is actually related to the average of the exponential of the field dual to $\phi$ (up to constants). This is because $\phi, n$ are canonically conjugate. Finally, $S_{\rm{top}}$ is merely the charge-$e$ boson correlation function. As a result it decays exponentially when the charge-$e$ spectrum is gapped and algebraically when the same is gapless. 

The Hamiltonian of Eqs. (\ref{ham2maj_2a}, \ref{ham2maj_2b}) can be mapped using the standard transfer-matrix method~\cite{Kogut1979} to a classical generalized XY model at finite temperature~\cite{Lee1985, Serna2017}. The resultant classical action has two interactions of the form:
\begin{equation}
{\cal A} = -\Delta_M\sum_{\langle i,j\rangle}\cos\Bigg(\frac{\theta_i-\theta_j}{2}\Bigg)-\Delta_J\sum_{\langle i,j\rangle}\cos\big(\theta_i-\theta_j\big),
\end{equation}
where $\Delta_{M,J}$ are the couplings of the classical model corresponding to $E_{M,J}$ and the $\{\theta_i\}$-s are the (classical) $U(1)$ degrees of freedom. The phase-diagram of this model has been analyzed in Ref.~\onlinecite{Serna2017} using a Monte Carlo method. As we will show below, our DMRG analysis of the quantum model is entirely compatible with the Monte Carlo results. 

Now, we discuss the anticipated phases and the phase-transitions of the JW-transformed model of Eqs.(\ref{ham2maj_2a}, \ref{ham2maj_2b}). The consequences for the original fermionic model can directly be  inferred from the following discussion. 

\subsection{Anticipated Phase-diagram of the MBH Model}
\label{phasediag}
We start by discussing some limiting cases of the model. Consider the case $E_M = 0$. Then, the spins decouple from the rotors and play no role in the phase-transitions of the model. The phase-diagram is simply that of the Bose-Hubbard model in the limit of high-occupancy of the boson sites~\cite{Fisher1989, Fazio2001, Sachdev2011} [see Eqs.~\eqref{ham2maj_2a},~\eqref{ham2maj_2b}]. The phase-diagram consists of lobes within which the system is a MI (see Fig.~\ref{pd1}). In this phase, the occupation of each island is pinned to an integer value. Both charge-$e$ and charge-$2e$ spectra are gapped. Thus, $S_b(i,j)$ and $S_{\rm{top}}(i,j)$ decay exponentially with $|i-j|$. On the other hand, $S_{\rm{triv}}(i,j)$ is a pinned to a constant (nonzero) value. Outside the lobes, the system is a charge-$2e$ LL. In the latter phase, the charge-$e$ spectra remains gapped, while the charge-$2e$ spectra is gapless. In the scaling limit, the LL phase is described by a conformal field theory (CFT) with central charge $c=1$, whose (euclidean) action is given by 
\begin{equation}
\label{lutt2e}
{\cal S} = \frac{1}{2\pi K_2}\int\ d^2x (\partial_\mu\phi)^2,
\end{equation}
where repeated summation with euclidean metric is assumed and we have set the velocity of the plasmons to be unity. From this action, it straightforwardly follows that the bosonic operators ${b}_i$ show a characteristic power-law decay with an exponent given by the LL-parameter, $K_2$ (see, for instance, Chap. 9 of Ref. ~\onlinecite{diFrancesco1997})
\begin{equation}
\label{lutt2}
S_b\sim \frac{1}{|i-j|^{K_{2}/2}}.
\end{equation}
In this phase, $S_{\rm{top}}$ is exponentially decaying, while $S_{\rm{triv}}$ is a again pinned to a constant (nonzero) value. The transition between the MI and the charge-$2e$ LL occurs at constant density through the tips of the lobes. The latter transition is of KT type.~\cite{Glazman1997, Kuhner2000} At the transition point, $K_2 = 1/2$. The transition occurring through the sides of the lobes occur with a change in the density. For this transition, $K_2 = 1$.

Next, consider the limit: $E_J=0$. Replacing $b$ by $\tilde{b}$ in the above argument, one can again conclude that the phase-diagram consists of lobes where the system is a MI and a charge-$e$ LL elsewhere. In the MI phase, the occupation of each island is pinned to half-integer values. This phase again has a gapped bulk spectrum and thus, the correlation functions $S_b$ and $S_{\rm{top}}$ decay exponentially, while $S_{\rm{triv}}$ is a constant. In the charge-$e$ LL phase, the charge-$e$ spectrum is gapless, which implies the same for the charge-$2e$ spectrum. In the scaling limit, the system is again described by Eq.~\eqref{lutt2e}. Now, all correlation functions decay algebraically:
\begin{align}
\label{lutt1}
S_{\rm{triv}}\sim \frac{1}{|i-j|^{K'/2}},\ S_{\rm{top}}\sim \frac{1}{|i-j|^{K_1/2}}
\end{align}
and the behavior of $S_b$ is given in Eq.~\eqref{lutt2}. From dimensional analysis, it follows that $K_1 = K_2/4$ and $K' = 1/K_1$. The nature of the transitions is the same as in the previous case and $K_2$ should be replaced by $K_1$. Furthermore, since the binding energy of a $4\pi$ vortex-antivortex pair is four times larger than that of a $2\pi$ vortex-antivortex pair (see, for instance, Chap. 9 of Ref.~\onlinecite{Chaikin2000}), the critical point occurs at about four-times smaller energy:
\begin{equation}
\label{est}
(E_M/E_C)_{\rm{crit}}\sim\frac{1}{4}(E_J/E_C)_{\rm{crit}}
\end{equation}

Finally, consider the limit when both $E_J, E_M$ are nonzero. From the above considerations, the system has three phases: MI, charge-$2e$ LL and the charge-$e$ LL. Tuning these tunneling rates with respect to $E_C$ leads to a phase-transition between the different phases. We show that the transition between the charge-$2e$ LL phase and the MI phase is always of KT type. The transition between the two LL phases (both $c=1$ CFT-s) is an Ising type transition. Physically, this Ising degree of freedom emerges to transition the system from a $2\pi$-periodic to a $4\pi$ periodic phase. At this transition, both the $U(1)$ and the Ising degrees of freedom undergo long-wavelength fluctuations. As a result, the total effective central charge at the transition is 3/2. Interestingly, the Ising transition line continues beyond the phase-boundary between the charge-$2e$ LL and the MI phases. In this region, the system directly undergoes a transition from the MI to the charge-$e$ LL phase. Finally, this Ising line turns into a KT line. We note that similar results were obtained with Monte Carlo method for the corresponding classical model~\cite{Serna2017}. 

Now, we present the DMRG results for the MBH model. 

\section{DMRG Results}
\label{dmrg}
The DMRG simulations were performed using the TeNPy package~\cite{Hauschild2018}. We simulated the JW transformed model given in Eqs.~(\ref{ham2maj_2a}-\ref{ham2maj_2b}). Techinical details regarding the DMRG simulations are provided in Appendix~\ref{extra_DMRG}.
\begin{figure}
\includegraphics[width = 0.5\textwidth]{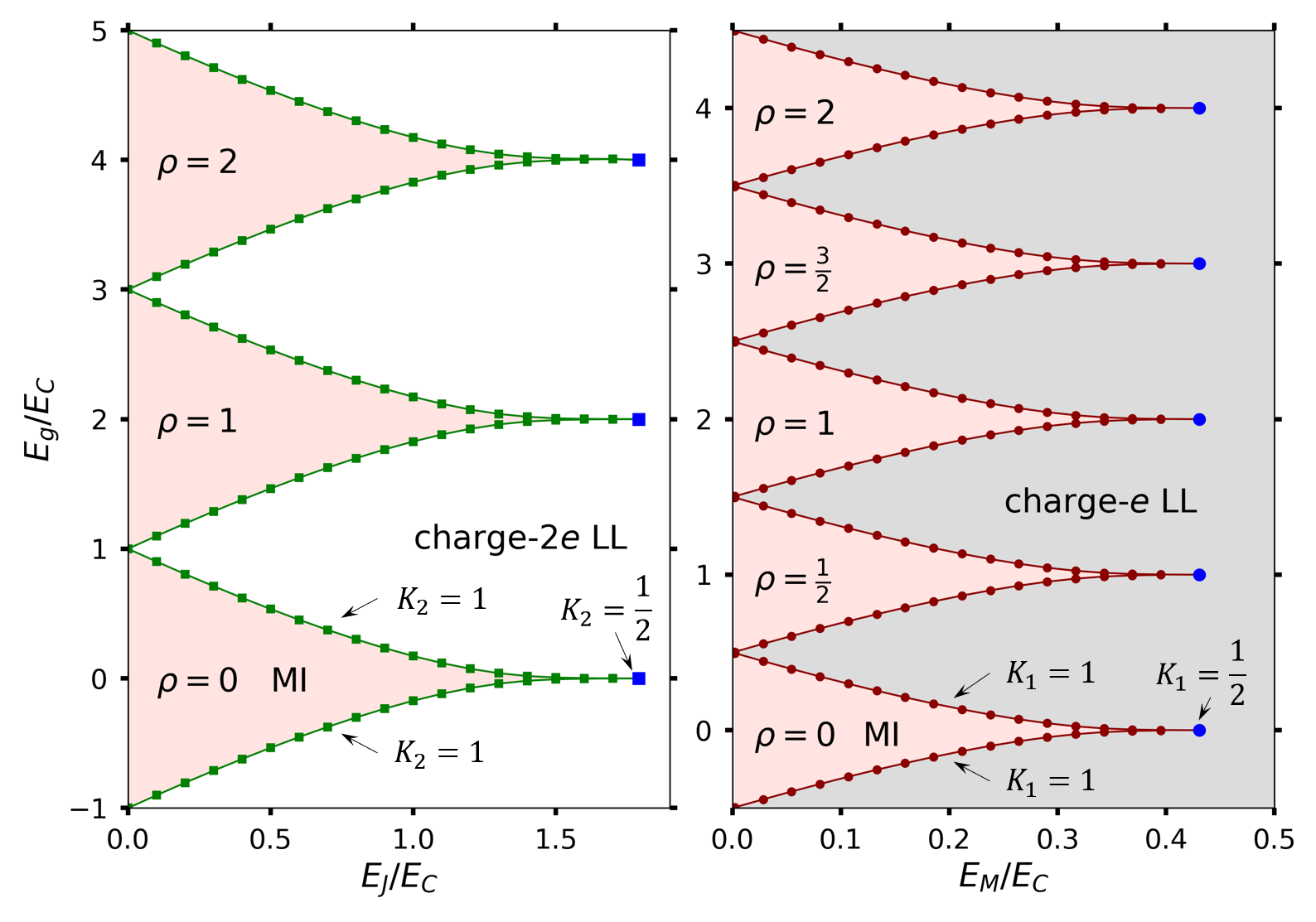}
\caption{\label{pd1} Phase-diagram of the MBH model for $E_M = 0$ (left panel) and $E_J = 0$ (right panel) obtained using DMRG. The system is in a MI phase within the lobes, while outside the system is in a charge-$2e$ (charge-$e$) LL phase, as shown in white (gray) in the left (right) panel. In the MI phase, the average occupation on each island is pinned to integer (half-integer) value for $E_M (E_J) = 0$ as shown in the left (right) panel. The phase-transition at constant density is of KT type, where the LL-parameter $K_{2(1)} = 1/2$ for $E_{M(J)} = 0$, as shown by the blue dot~(square). The phase-transition along the sides of the lobe is accompanied by a change in the density. The LL-parameter for this phase-transition is given by $K_{2(1)} = 1$ for $E_{M(J)} = 0$~[the maroon~(green) dots~(squares)]. The characteristics of the different lobes on each of the panels are identical. We have only annotated the bottom-most lobes in the figure for brevity. }
\end{figure}

Fig.~\ref{pd1} shows the phase-diagram of the system for the two limiting cases of $E_M=0$ (left panel) and $E_J=0$ (right panel) respectively obtained using DMRG calculations. The phase-diagram consists of the Mott lobes within which the system is in an insulating phase, where the average occupation number per island is pinned to an integer (half-integer) value for $E_{M(J)} = 0$. As $E_{J(M)}$ increased, the system undergoes a transition from the MI phase to a charge-$2e(e)$-LL phase. Across the sides of the lobe, the phase-transition is accompanied by a change in the density. The location of the sides are obtained by computing the cost of adding an extra particle/hole and setting this to be equal to the gate voltage. To that end, we simulated system sizes: $L = 32, 64,$ $96$ and $128$ and used finite-size scaling to obtain the gate voltages at the boundaries of the Mott lobes. Note that the single-particle/hole excitation gap scales linearly with inverse of the system size as in the conventional 1D Bose-Hubbard model.~\cite{Kuhner2000}  A maximum bond-dimension of $120$ was used to keep the truncation errors in the order of $10^{-8}$. The phase-transition through the tip of the lobe is at constant density and is of KT type. To improve precision, the location of this phase-transition is determined by calculating the Luttinger parameter $K_{2(1)}$ and checking when it crosses $1/2$. The Luttinger parameter was obtained by computing the correlation functions, $S_{b({\rm top})}$, given in Eqs.~\eqref{lutt2},~\eqref{lutt1} and fitting on a log-log scale. While the lobes were obtained by doing finite system simulations, these correlation functions were computed using infinite-DMRG~(iDMRG) to reduce finite-size effects (see Appendix~\ref{extra_DMRG_KT} for more details). The location of the KT transition points for the left and right panels of Fig.~\ref{pd1} are given by 
\begin{align}
(E_J/E_C)_{\rm{crit}} &= 1.780\pm 0.005,\nonumber\\ (E_M/E_C)_{\rm{crit}} &= 0.432\pm 0.002,
\end{align}
which satisfy the rough estimate given in Eq. \eqref{est}.

\begin{figure}
\includegraphics[width = 0.5\textwidth]{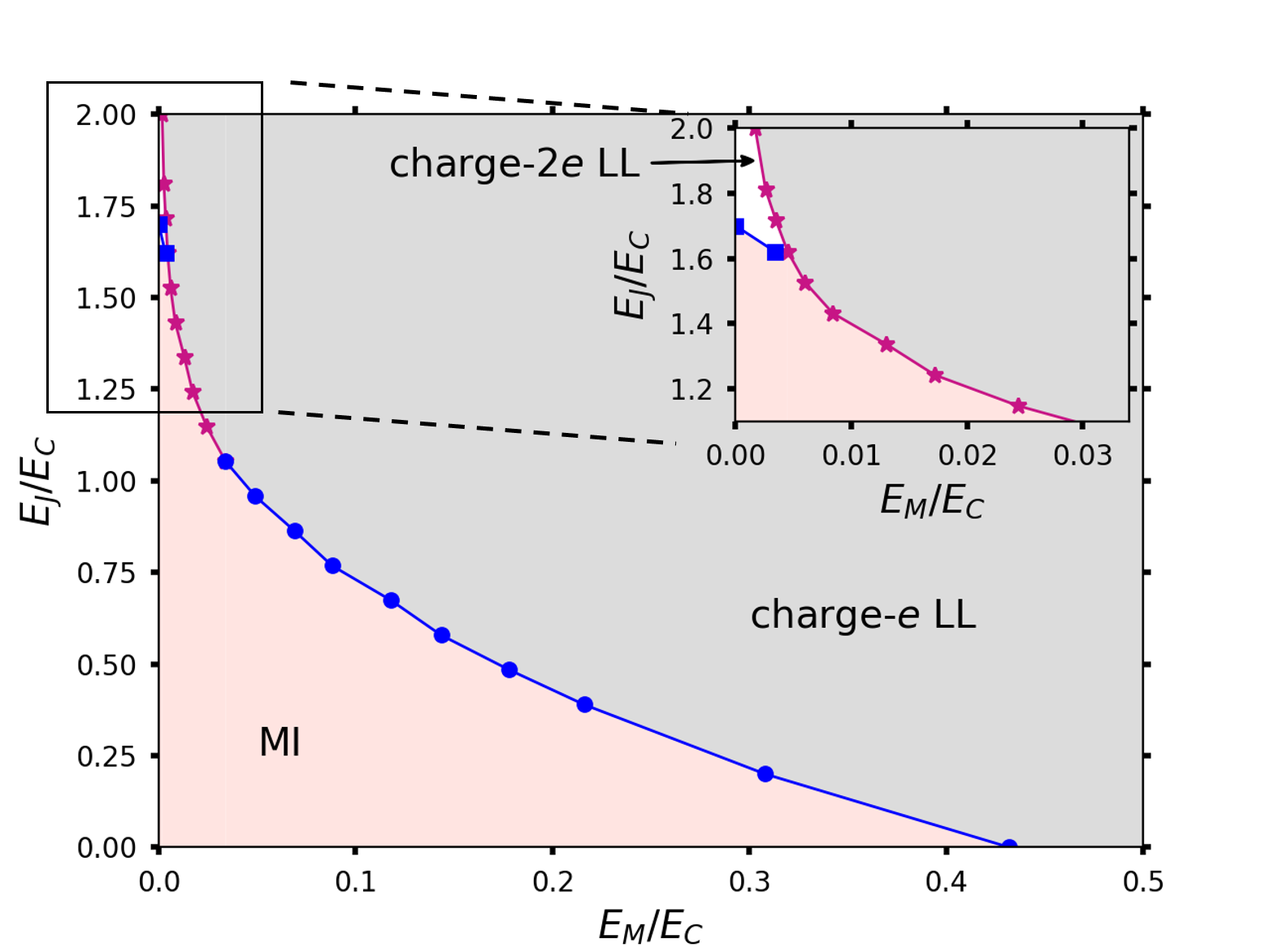}
\caption{\label{pd} The phase-diagram of the MBH model for nonzero $E_J/E_C, E_M/E_C$. We chose $E_g = \rho =0$. The charge-$2e$ LL phase and the MI phase are separated by a KT transition (blue squares), where $K_2 = 1/2$. This KT transition line meets an Ising transition line (violet-red stars), which separates the two LL phases. This Ising line continues beyond the KT line and separates part of the MI and the charge-$e$ LL phases. Finally, this line turns into a KT line (blue dots). At the latter KT transition, $K_1=1/2$. (Inset) Zoom in on the region where the three phases meet.}
\end{figure}

The phase-diagram for both $E_J, E_M\neq0$ is shown in Fig.~\ref{pd}. We only show the case $E_g/E_C =\rho =  0$ for the ease of presentation. We used a combination of finite and infinite DMRG techniques. The DMRG results validate the anticipated phase-diagram explained in Sec. \ref{phasediag}. For $E_M/E_C\ll1$, upon increasing $E_J/E_C$, the system transitions from the MI phase to the charge-$2e$ LL phase through a KT transition (blue squares). Along this transition line, $K_2=1/2$. Increasing $E_M/E_C$ while being in the charge-$2e$ LL phase causes the system to transition to the charge-$e$ LL phase through an Ising transition (violet-red stars). Upon lowering $E_J/E_C$, this Ising line continues beyond the phase-boundary of the charge-$2e$ LL and the MI phases and also separates the MI and the charge-$e$ LL phase. Further lowering $E_J/E_C$, this Ising line turns into a KT line that emerges from below (blue dots). Along this KT line, $K_1 = 1/2$. A zoom in on the region where the three phases meet is shown in the inset of Fig.~\ref{pd}. 

\begin{figure}[!ht]
\includegraphics[width = 0.5\textwidth]{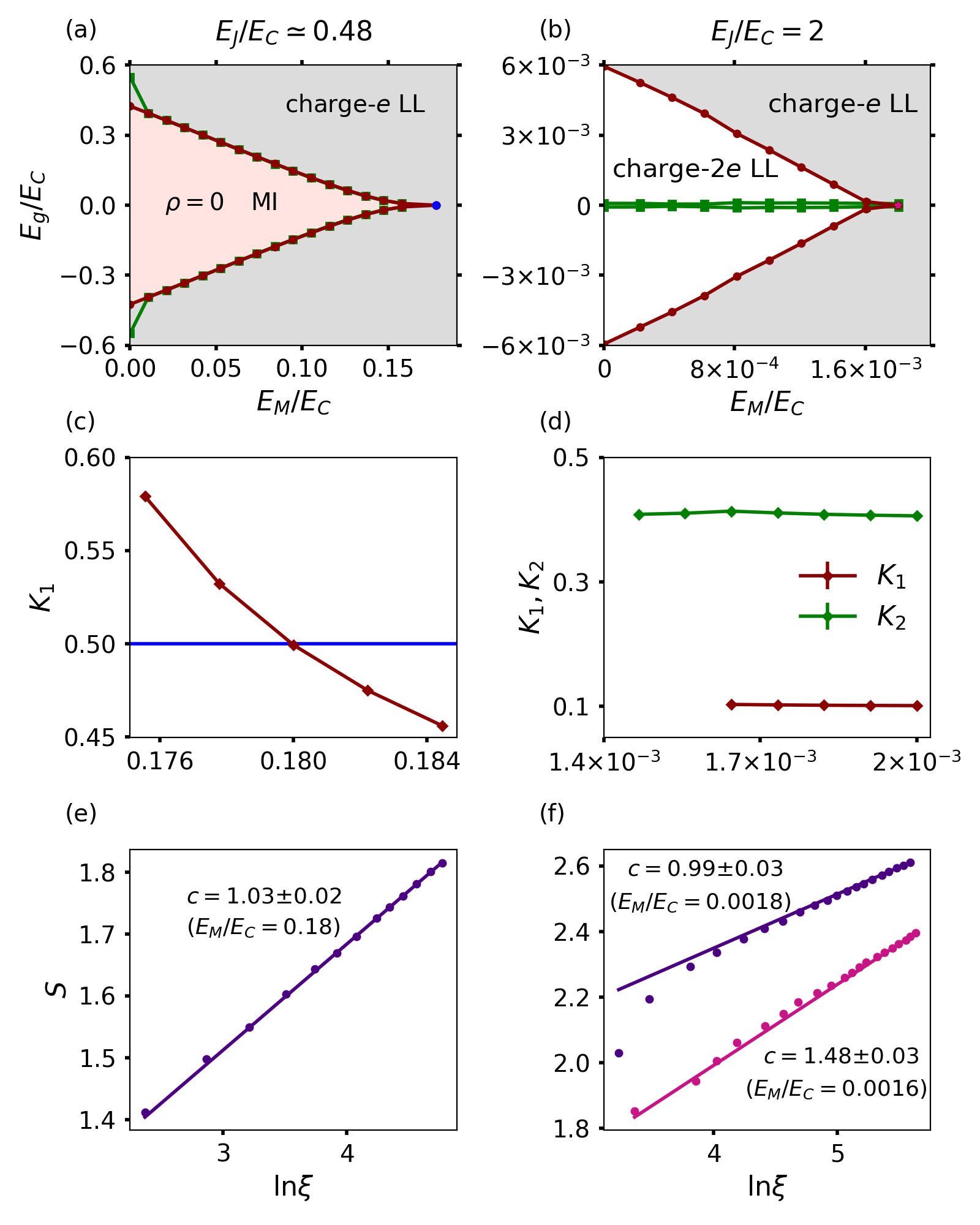}
\caption{\label{lobe_2} Qualitatively different phase-transitions the system undergoes to enter the charge-$e$ LL phase from the MI and the charge-$2e$ LL phase. We chose $\rho = 0$. (a,c,e) Transition from the MI to the charge-$e$ LL. We chose $E_J/E_C\simeq 0.48$. (a) The system is a MI inside the lobe and a charge-$e$ LL everywhere else. Both charge-$e$ (dark red dots) and charge-$2e$ (green squares) excitation gaps are computed to obtain the gate voltages at the boundary of the lobe. For any nonzero $E_M/E_C$, both the charge-$e$ and charge-$2e$ spectra are gapless simultaneously at the sides and the tip of the lobe. (c) Zoom in at the tip of the lobe (blue dot). The nature of the phase-transition is of KT type, where $K_1$ (dark red diamonds) $= 1/2$ (blue line). (e) Scaling of entanglement entropy ($S$) vs. log of correlation length ($\xi$) in the charge-$e$ LL phase used to extract the central charge ($c$). (b,d,f) Transition from the charge-$2e$ LL to the charge-$e$ LL. We chose $E_J/E_C=2$. (b) The system is a charge-$2e$ LL inside the white lobe, while being a charge-$e$ LL everywhere else. (d) Zoom in around the tip of the lobe. The Luttinger parameter $K_2$ is below 1/2, while $K_1$ abruptly drops from $>10$ in the charge-$2e$ LL phase (not shown) to below 1/2. (f) Scaling of $S$ vs.~$\ln\xi$ used to confirm the nature of the transition at the lobe-tip (in violet-red) contrasted with the same outside the lobe (in purple). The extra 1/2 in the central charge at the tip of lobe indicates the additional gapless Ising degree of freedom on top of the gapless $U(1)$.}
\end{figure}

Next, in Fig.~\ref{lobe_2}, we show the fundamentally different nature of the phase-transitions occuring as the system enters the charge-$e$ LL phase. For $E_J/E_C \simeq 0.48$ (left half of Fig.~\ref{lobe_2}), the system is a MI inside the lobe, where both charge-$e$ and charge-$2e$ spectra are gapped. For any nonzero $E_M/E_C\neq 0$, at the sides and the tip of the lobe, both the $e, 2e$ gaps close simultaneously. This is indicated by the overlap of the green squares and the dark red dots. The nature of the phase-transition across the tip is of KT type and is driven by binding of $4\pi$ vortices, when $K_1 = 1/2$ [Fig.~\ref{lobe_2}(c)]. In Fig.~\ref{lobe_2}(e), we show the scaling of entanglement entropy ($S$) vs. log of the correlation length ($\xi$) in the charge-$e$ LL phase. From the scaling, we extract a central charge of $1.03\pm0.02$. Now, consider the case when $E_J/E_C=2$ (right half of Fig.~\ref{lobe_2}). The $2e$ spectrum is already gapless for this parameter choice (the tiny gap visible between the green squares is of the order of $10^{-5}$ and is a finite size effect; we checked that there was a systematic decrease in this gap as larger system sizes were simulated). The charge-$e$ excitation gap closes as $E_M/E_C$ is increased from zero. Note that the charge-$e$ excitation gap in the charge-$2e$ LL phase can be as small as $10^{-4}$ [Fig.~\ref{lobe_2}(b)]. Thus, to obtain the charge-$2e$ LL lobe, we computed the charge-$e$ excitation gap for system sizes: 80, 128, 256, 320, 400 and used finite-size scaling to obtain the gap. Furthermore, we chose a maximum bond-dimension of $400$ to keep truncation errors below $10^{-8}$. The charge-$e$ gap closes at the sides and the tip of the lobe. The transition at the tip of the lobe is of Ising type. While in the two LL phases on either side of the transition, only a  $U(1)$ degree of freedom is gapless, at the transition, an emergent Ising degree of freedom becomes gapless. As a result, the effective central charge rises from 1 on either side of the transition to 3/2 at the transition. To capture this, we compute the central charge as we tune $E_M/E_C$ across the transition using iDMRG [Fig.~\ref{lobe_2}(f)]. From the scaling of $S$ vs.~$\xi$, we extract a central charge of $1.48\pm0.03$ at the transition occurring at $E_M/E_C\simeq0.0016$. For comparison, we have also shown the $S$ vs. $\ln\xi$ scaling for $E_M/E_C\simeq0.0018$, when the system is in the charge-$e$ LL phase. At the latter point, the central charge turns out to be $0.99\pm0.03$. To verify that the transition is not of KT type, we computed the Luttinger parameters, $K_1, K_2$, across the transition. As shown in Fig.~\ref{lobe_2}(d), $K_2$ is below 1/2 for this range of parameters. The value of $K_1$ drops abruptly to below 1/2 from $>10$ (these large values are not shown in figure to focus on those around 1/2)  across the transition, in contrast to slowly varying across 1/2.

While the numerics was done only for the tip of the lobe, the nature of the transition at the sides of the lobe is presumably also Ising, based on purely symmetry considerations. The complexity of the numerical simulations to capture the Ising transition increases substantially compared to the KT transition; more details are provided in Appendix \ref{DMRG_Ising}. We saw similar behavior for the remaining points of the Ising transition and did not observe any first-order transition. To rule out first order transitions, we also computed the fidelity susceptibility~\cite{Gu2010}. We did not see the fidelity of the ground state for two successive parameter values of $E_M/E_C$ to dip to zero, as would be the case for a first-order transition. Our results are compatible with the Monte Carlo results for the generalized XY model of Ref.~\onlinecite{Serna2017}, who also observe only a continuous Ising transition.

% %The $2e$ spectrum is gapless everywhere, while the $e$ spectrum is gapless only in the charge-$e$ LL. 
\begin{figure}
\includegraphics[width = 0.5\textwidth]{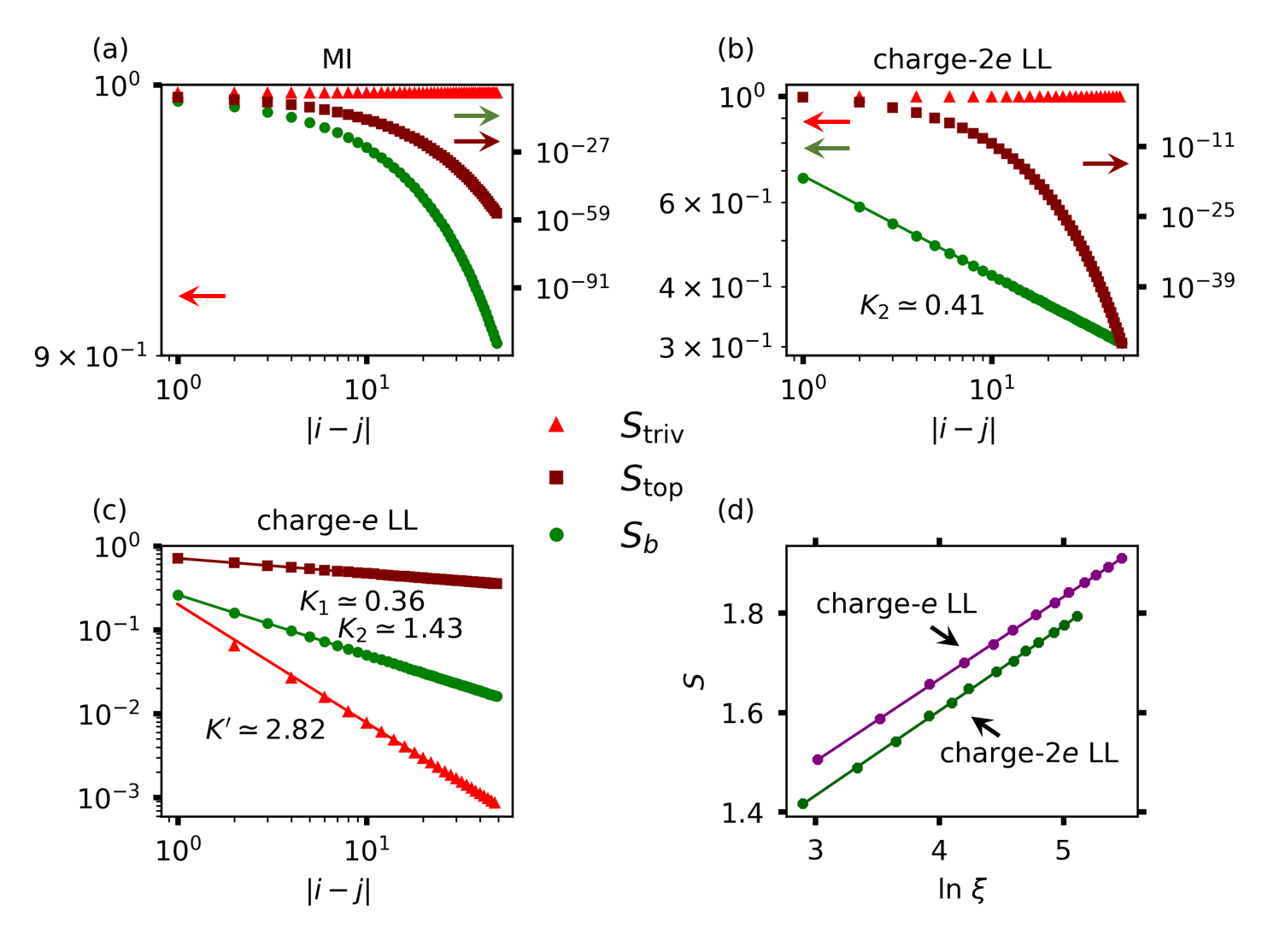}
\caption{\label{threephases} (a-c) Characteristic decay of $S_b$, $S_{\rm{triv}}$ and $S_{\rm{top}}$ in the MI, charge-$2e$ LL and the charge-$e$ LL phases (see Table \ref{tab1}) using iDMRG technique. We chose $E_J/E_C =0, E_M/E_C = 0.02$ for the MI phase, $E_J/E_C = 0$, $E_M/E_C = 0.6$ for the charge-$e$ LL phase and $E_J/E_C = 1$, $E_M/E_C = 2\times10^{-4}$ in the charge-$2e$ LL phase. (a) In the MI phase, $S_{\rm{triv}}$ is constant and nonzero, while both $S_{\rm{top}}, S_b$ are exponentially decaying. (b) In the charge-$2e$ LL phase, $S_{\rm{triv}}$ is constant and nonzero, while $S_{\rm{top}}$ is exponentially decaying. On the other hand, $S_b$ decays algebraically. The extracted Luttinger parameter is: $K_2 \simeq 0.41$. (c) In the charge-$e$ LL phase, all the three correlators decay algebraically. The extracted Luttinger parameters are: $K_1 \simeq 0.36$, $K_2\simeq1.43$ and $K' \simeq 2.82$. (d) Variation of the entanglement entropy ($S$) with the log of the correlation length ($\xi$) in the charge-$e, 2e$ LL phases. Both phases show the characteristic linear dependence with the extracted central charge $c=1$ (with precision to the second decimal place). }
\end{figure}
After the different phase-transitions that separate the different phases, we turn to characterize the latter. In particular, we compute the three order parameters, $S_b, S_{\rm{triv}}$ and $S_{\rm{top}}$ in the three different phases, shown in Fig.~\ref{threephases}. We chose $E_J/E_C =0, E_M/E_C = 0.02$ for the MI phase, $E_J/E_C = 0$, $E_M/E_C = 0.6$ for the charge-$e$ LL phase and $E_J/E_C = 1$, $E_M/E_C = 2\times10^{-4}$ in the charge-$2e$ LL phase. As explained in Sec. \ref{model} (see Table \ref{tab1}), these behaviors can be inferred from the gapless/gapped nature of the charge-$e, 2e$ spectra in the different phases. The trivial correlator, $S_{\rm{triv}}$, is nonzero in the MI and the charge-$2e$ LL phase, while it decays algebraically in the charge-$e$ LL phase. On the other hand, the topological correlator, $S_{\rm{top}}$, is exponentially decaying in the MI and the charge-$2e$ LL phases, while it decays algebraically in the charge-$e$ LL phase. Finally, $S_b$ decays algebraically in both the LL phases, while it decays exponentially in the MI phase. 

\begin{figure}
\includegraphics[width = 0.5\textwidth]{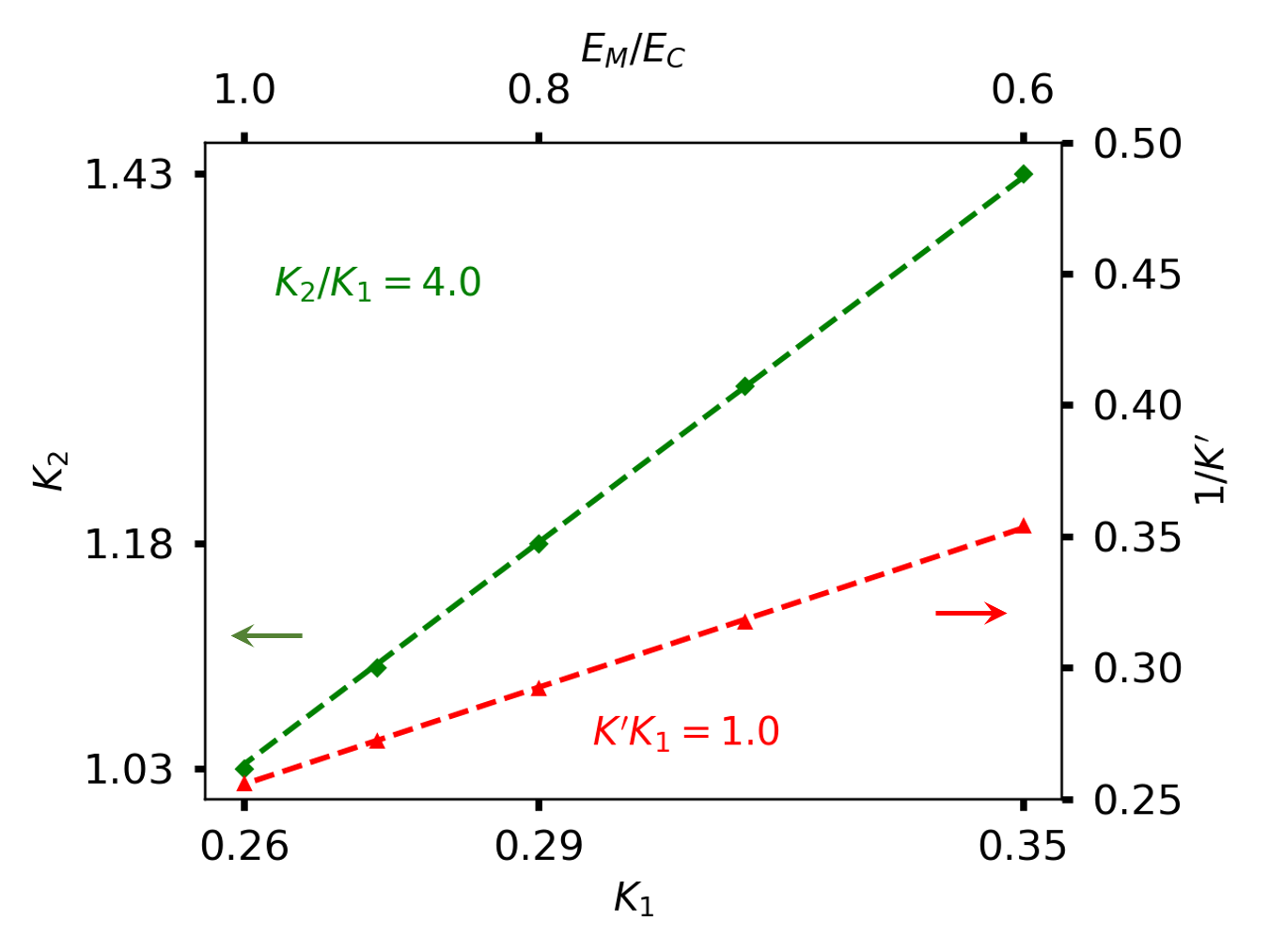}
\caption{\label{luttcentch} Verification of the relative magnitude of the Luttinger parameters deep in the charge-$e$ LL phase for $E_J = \rho = 0$ using iDMRG technique. The values of the $E_M/E_C$ are shown on the top x-axis.  The green diamonds (red triangles) are the evaluated values of $K_2$ ($1/K^\prime$) as a function of the evaluated values of $K_1$. The dashed lines are linear fits, which confirm the relations between the different Luttinger parameters: $K_2/K_1 = 4$, $K^\prime = 1/K_1$. }
\end{figure}

As explained in Sec. \ref{phasediag}, the rates of the algebraic decay of the three correlators in the charge-$e$ LL phase, given by the three Luttinger parameters, $K_1, K_2, K^\prime$, are related to each other. Fig.~\ref{luttcentch} shows the verification of the relationships using DMRG.  We chose $E_J/E_C = 0$ and varied $E_M/E_C$. Fitting the obtained values of $K_2$ and $1/K'$ with $K_1$, we obtain the relations: $K_2 = 4K_1$ and $K'= 1/K_1$, which validate the predictions of Sec.~\ref{phasediag}.

\section{The nature of correlations in the charge-$e$ LL phase}
\label{topornot}
We have analyzed the phase-diagram of the MBH model in the previous section. In this section, we discuss the nature of the correlations in the charge-$e$ LL phase and how it relates to the topological phase of the Kitaev chain~\cite{Kitaev2001}. The Hamiltonian of the Kitaev chain can obtained by adding a term of the form $E_J^b\sum_{i=1}^L\cos\phi_i$ and considering the limit $E_J^b\gg E_J, E_M, E_C$. Physically, this would correspond to adding a large bulk superconductor under the mesoscopic islands. This bulk superconductor breaks the $U(1)$ symmetry of the model discussed in this work, reducing it to $\mathbb{Z}_2$. The Kitaev chain has two phases: the gapped, trivial phase and the gapped, topological phase, the latter characterized by the existence of topologically protected zero-energy edge-modes. The nonlocal string correlation function, $S_{\rm{triv}}(S_{\rm{top}})$, is zero (nonzero) in the topological phase and the other way around in the trivial phase.  

In contrast, for the MBH model, both $S_{\rm{top}}$ and $S_{\rm{triv}}$ decay algebraically in the charge-$e$ LL phase, the rate of decay being $K_1$ and $1/K_1$. For the Hamiltonian given by Eq.~\eqref{ham2maj}, the $K_1$ is always $\leq1$. This is best seen from the JW-transformed model of Eqs.~(\ref{ham2maj_2a}, \ref{ham2maj_2b}). As is well-known, in the presence of only onsite-repulsion, the LL phase of the Bose-Hubbard model always has Luttinger parameter smaller than unity, {\it i.e.}, the LL is attractive~\cite{Giamarchi2003}. If computed inside the MI phase, the Luttinger parameter is larger than unity, but in this phase, the Gaussian action of Eq.~\eqref{lutt2e} ceases to be a good description of the system. Thus, for the model analyzed, the decay of $S_{\rm{top}}$ is always slower than that of $S_{\rm{triv}}$. 

However, this behavior can be altered by adding {\it local} perturbations. Adding nearest neighbor repulsion of the form $V\sum_{i=1}^{L-1}n_in_{i+1}$ can cause the system to transition to a repulsive LL phase, where the Luttinger parameter $K_1>1$.~\cite{Glazman1997, Kuhner2000} For the repulsive LL, $S_{\rm{triv}}$ decays more slowly than $S_{\rm{top}}$. The attractive and repulsive LL-s are separated by a phase-transition only in the following sense. In the presence of an impurity, the transport signatures are different~\cite{Schmid1983, Caldeira_Leggett_1983,Kane1992,Kane1992a}. Consider the case when there is an impurity in the LL. In the superconducting context, this can be done by adding a weak-link junction between two halves of the charge-$e$ LL. Then, for an attractive (repulsive) LL, the impurity disappears (cuts the LL in half).~\cite{Schmid1983,Glazman1997} 

The decay of $S_{\rm{triv}}$~vs.~$S_{\rm{top}}$ in the attractive and repulsive charge-$e$ LL phases indicates the proximity of these phases to a topological phase. Upon inclusion of the $U(1)$-symmetry breaking perturbation of the form $E_J^b\sum_{i=1}^L\cos\phi_i$, the attractive (repulsive) charge-$e$ LL is transformed into a topological (trivial) phase. We emphasize that the charge-$e$ LL is {\it not} a ``topological'' or ``quasi-topological'' phase~\cite{Bonderson2013}, but is in the proximity of a topological phase. 

There are also no zero-energy edge-modes in the charge-$e$ LL phase. This is again a consequence of the fact that there is no way to break the parity symmetry  $\prod_{i=1}^LZ_i$ without breaking the $U(1)$ symmetry. We checked this using DMRG. For finite systems with open boundary conditions, we see that the energy required to create a fermionic excitation or a bosonic one is the same. As a result, the levels are split in the same way as the bulk excitation spectrum $\sim 1/L$ (this $1/L$ scaling was used to get the Mott lobes in the first place). We point out that this result is in contrast to those obtained by bosonization calculations of a proximitized superconductor~\cite{Fidkowski2011}. While the bosonization calculations predict a fermionic gap despite the presence of gapless bosonic excitations, our results indicate gapless bosonic and fermionic excitations. We note that this behavior for this model was already predicted in Ref. ~\onlinecite{Xu2010}. The source of this difference is due to the fact that we treat a fundamentally different model, that of an inherent p-wave superconductor with phase-fluctuations, instead of a proximitized nanowire. It will be interesting to verify the results of the bosonization computation using DMRG as well and we hope to return to this problem in the future.

\section{Conclusion and Perspectives}
\label{concl}
To summarize, we have numerically investigated the fate of the topological phase of a 1D spinless, p-wave superconductor at zero temperature in the presence of phase-fluctuations using the DMRG technique. We consider the phase-fluctuations that occur when the p-wave superconductor is in contact with a 1D s-wave superconductor, as opposed to a 3D bulk s-wave superconductor~\cite{Kitaev2001}. The discretized model used is the MBH model, which describes the interactions of a 1D array of MZM-s on mesoscopic superconducting islands. The chain of mesoscopic islands models the quasi-long-range order of the 1D s-wave superconductor with power-law phase-correlations. Cooper-pair and MZM-assisted single-electron tunneling, together with the finite charging energy of the mesoscopic islands, lead to a rich phase-diagram of the model. We show that the system can be in a MI, a charge-$2e$ LL or a charge-$e$ LL phase. The first two phases are separated by a KT transition. Upon tuning of the single-electron tunneling rate, the system  undergoes quantum phase-transitions to a charge-$e$ LL, which can be either KT or Ising type. This charge-$e$ LL phase is in the proximity of the topological phase of Ref.~\onlinecite{Kitaev2001}. Nonlocal string correlation functions, suitably generalized to include the effect of phase-fluctuations, show the fundamental difference between the nature of ordering that occurs due to the presence of phase-fluctuations. We show that the relevant nonlocal string correlation function decays algebraically in the charge-$e$ LL phase due to the presence of gapless charge-$e$ excitations. This should be contrasted to the topological phase of Ref.~\onlinecite{Kitaev2001} where it is constant and nonzero asymptotically. Furthermore, we show that the MBH model can be mapped to other known models of quantum and classical statistical mechanics. By performing a JW transform, together with additional gauge transformations, we map the MBH model to a generalized Bose-Hubbard model with two types of hopping. Subsequently, standard transfer-matrix mapping shows that the critical properties of the MBH model are identical to that of a generalized classical XY model at finite temperature.~\cite{Lee1985, Serna2017} Our DMRG results are compatible with the Monte Carlo computations of the classical model. 

Before concluding, we discuss some generalizations of the MBH model which show interesting properties and are of current research interest. First, consider $2\alpha$ ($\alpha>1$) instead of two MZM-s on each island. This corresponds to a stacking of Kitaev chains~\cite{Pollmann2011, Fidkowski2011a, Verresen2017} in the presence of phase-fluctuations. By appropriately choosing the JW transformation contour, these models can be reduced to the case analyzed in this work. We do this explicitly for the case $\alpha = 2$ in Appendix \ref{2maj}. Due to the presence of an extensive number of conserved quantities, the Hilbert space of the model splits into sectors, with only certain sectors of the Hilbert space having access to the whole phase-diagram. Second, the case of four MZM-s on each mesoscopic islands in 2D shows an even more interesting phase-diagram~\cite{Xu2010, Terhal2012, Roy2017}. There are again three phases which are the 2D counterparts of the 1D case analyzed in this work: MI, topological superconductor and ordinary superconductor. Furthermore, the MI and the ordinary superconducting phases exhibit toric code ordering, which is one of the most promising platforms for quantum computation. The field theory calculations for the phase-diagram of this model and the associated charge-response have been done~\cite{Roy2017, Roy2018a}. The field theory computations predict a couple of tricritical points and first-order transitions between the Ising and the KT transitions that separate the topological superconductor from the ordinary superconductor and the MI. This should be contrasted with results obtained in this work where the Ising transition smoothly merges with the KT line. It will be interesting to see if the numerical analysis of the 2D model supports the field-theory calculations. We hope to report on this in the future.  

\section*{Acknowledgments}
We gratefully acknowledge discussions with Ruben Verresen and Hubert Saleur. A.R. acknowledges the support of the Alexander von Humboldt foundation. We acknowledge the support of the DFG Research Unit FOR 1807 through grants no. PO 1370/2-1, TRR80 through grant no. 107745057, the European Research Council (ERC) under the European Union’s Horizon 2020 research and innovation program (grant agreement no. 771537), and the DFG under Germany’s Excellence Strategy - EXC-2111 - 390814868.
\appendix

\section{Generalization to four MZM-s on each superconducting island}
\label{2maj}
\begin{figure}
\includegraphics[width = 0.5\textwidth]{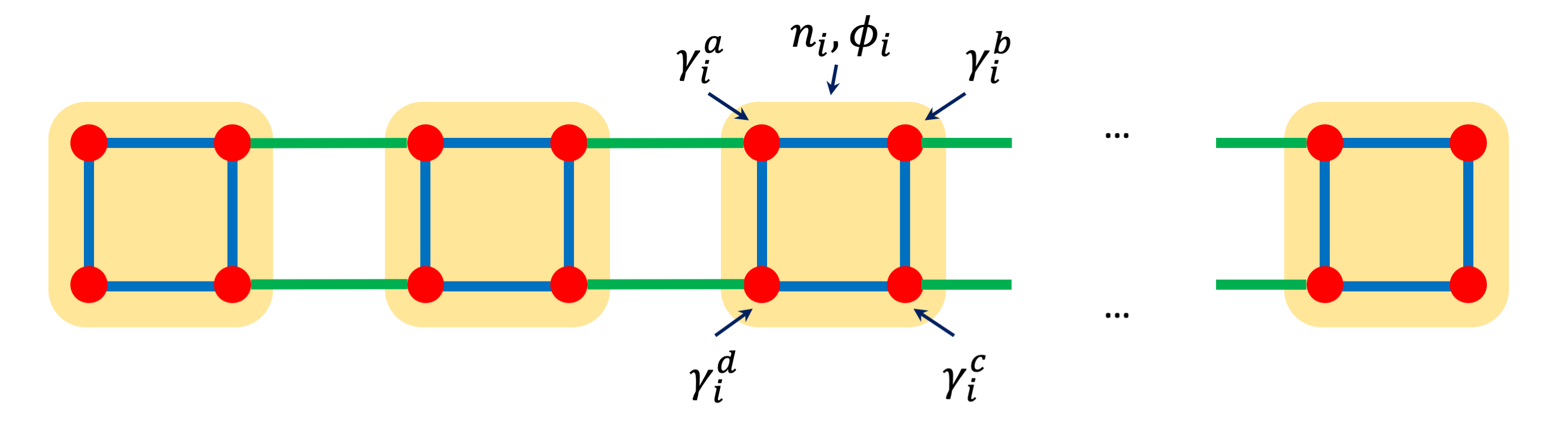}
\caption{\label{1darray2} Schematic of a 1D array of mesoscopic superconducting islands (denoted by yellow boxes) with MZM-s (denoted by red dots). In addition to the rotor degrees of freedom, $\phi_i, n_i$, each island has four MZM-s, denoted by $\gamma_i^a$, $\gamma_i^b$, $\gamma_i^c$ and $\gamma_i^d$. }
\end{figure}

In this section, we generalize the above model to the case when four MZM-s are present on the same mesoscopic island (see Fig.~\ref{1darray2}). The Hamiltonian describing the system is given by $H = H_C + H_J + H_M + H_g$, where $H_C, H_J$ and $H_g$ are still given by Eqs.~\eqref{ham2maj_a}, ~\eqref{ham2maj_c}. The MZM assisted single-electron tunneling Hamiltonian is now given by
\begin{align}
\label{ham4maj_b}
H_M &= E_M\sum_{i=1}^{L-1}\big(i\gamma_i^b\gamma_{i+1}^a+i\gamma_i^c\gamma_{i+1}^d\big)\cos\frac{\phi_i-\phi_{i+1}}{2}.
\end{align}
The constraint on the physical wavefunction is still given by Eq.~\eqref{parconstr}, but now the parity operator on the $i^{\rm{th}}$ island is given by $P_i=-\gamma_i^a\gamma_i^b\gamma_i^c\gamma_i^d$. 

As before, we again map the fermions to spins using a JW transformation. We take the JW string such that first the fermions on the top-row ($\gamma_i^a, \gamma_i^b$) are traversed from left to right and then those on the bottom row ($\gamma_i^c, \gamma_i^d$) are traversed right to left.\footnote{A different choice of the Jordan-Wigner string entails traversing all four MZM-s on each island in the order $a,b,d,c$ and the islands are traversed from left to right. This leads to the cluster model~\cite{Verresen2017}. However, not all the symmetries are easily apparent with this choice.} Explicitly, the mapping is given by
\begin{align}
\gamma_i^a &= \Big(\prod_{j<i}Z_j\Big)X_i, \ \gamma_i^b = \Big(\prod_{j<i}Z_j\Big)Y_i,\nonumber \\
\gamma_i^c &= \Big(\prod_{j=1}^L Z_j\Big)\Big(\prod_{k = L}^{i+1}\bar{Z}_k\Big)\bar{X}_i,\nonumber \\ \gamma_i^d &= \Big(\prod_{j=1}^L Z_j\Big)\Big(\prod_{k = L}^{i+1}\bar{Z}_k\Big)\bar{Y}_i,
\end{align}
where we have used $X(\bar{X}), Y(\bar{Y})$ and $Z(\bar{Z})$ to denote the spins on the upper (lower) row. The parity operator is transformed to $P_i = Z_i\bar{Z}_i$, while Eq.~\eqref{ham4maj_b} is transformed to
\begin{align}
H_M &= -E_M\sum_{i=1}^{L-1}\big(X_iX_{i+1}-\bar{X}_i\bar{X}_{i+1}\big)\cos\frac{\phi_i-\phi_{i+1}}{2}\nonumber.
\end{align}
It is easy to see that $\bar{X}_iX_i$, $i=1,\ldots,L$ commutes with the Hamiltonian and is thus, conserved.~\cite{Roy2018a} To make this symmetry more explicit, we rewrite the operators for the two spins on each island in a Bell-basis~\cite{Terhal2012}, described below. We define a sign qubit ($s$) and a target qubit ($t$) on each island, whose joined state is given by $|\psi^{s,t}\rangle \equiv |s,t\rangle$: 
\begin{align}
|s=0,t=0\rangle &= \frac{1}{\sqrt{2}}(|00\rangle + |11\rangle),\nonumber\\ |s=0,t=1\rangle &= \frac{1}{\sqrt{2}}(|01\rangle + |10\rangle), \nonumber\\
|s=1,t=0\rangle &= \frac{1}{\sqrt{2}}(|00\rangle - |11\rangle),\nonumber\\ |s=1,t=1\rangle &= \frac{1}{\sqrt{2}}(|01\rangle - |10\rangle),
\end{align}
where on the right-hand side of the above equations, the first (second) position in the kets refer to the spin on the upper (lower) row on each island. 
Thus, $s$ is the sign bit and the $t$ is the two-qubit parity bit of information in the superposition. In this basis, the operators $X_i, \bar{X}_i$, $Z_i, \bar{Z}_i$ get mapped to:
\begin{align}
X_i|s,t\rangle = X_i^tZ_i^s|s,t\rangle, \ \bar{X}_i|s,t\rangle = X_i^t|s,t\rangle,\nonumber\\
Z_i|s,t\rangle = X_i^s|s,t\rangle, \ \bar{Z}_i|s,t\rangle = Z_i^tX_i^s|s,t\rangle,
\end{align}
where $X_i^{s(t)}, Z_i^{s(t)}$ are the Pauli operators for the sign (target) qubits. Under this mapping, the conserved quantities $X_i\bar{X}_i$ get mapped to $Z_i^s$. The Hamiltonian can be written in this Bell-basis, after applying the unitary transformation of Eq.~\eqref{unitary}. After some algebra, we arrive at 
\begin{align}
\label{ham4maj_2a}
H_C &= E_C\sum_{i=1}^L \tilde{n}_i^2,\ H_J = -\frac{E_J}{2} \sum_{i=1}^{L-1}(b_{i}b_{i+1}^\dagger + {\rm{H.c}}),\\\label{ham4maj_2b}
H_M &= -\frac{E_M}{2}\sum_{i=1}^{L-1}(Z_i^sZ_{i+1}^s-1)(\tilde{b}_{i}\tilde{b}_{i+1}^\dagger + {\rm{H.c}}),\\  H_g &= -E_g\sum_{i=1}^L \tilde{n}_i.
\end{align}
Here, $\tilde{n}_i = n_i + (1+Z_i^t)/4$ and $\tilde{b}_i =  {b}_i\sigma_{i,t}^+ + \sigma_{i,t}^-$, where $\sigma_{i,t}^\pm$ are the operators of the $i^{\rm{th}}$ target qubit. Note that only the conserved $Z_i^s$ operators of the sign qubits are present in the Hamiltonian, while $\tilde{b}_i$-s are built out of only the target qubit operators $\sigma_{i,t}^\pm$ and the Cooper-pair operators $b_i$-s. For $Z_i^sZ_{i+1}^s=1$, the single-electron tunneling term disappears and the system can be either in the MI or the charge-$2e$ LL phase. On the other hand, for $Z_i^sZ_{i+1}^s = -1$, the system can be in either MI, charge-$e$ LL or the charge-$2e$ LL phase. As before, time-reversal symmetry and the $U(1)$ symmetry corresponding to the conservation of $\tilde{n}_i$ are also present for this model. 

Thus, the phase-diagram of this model is qualitatively similar to the MBH model with only two MZM-s on each island. The difference is that only certain sectors of the Hilbert space have access to the entire phase-diagram. While we do the analysis for four MZM-s, a similar analysis can presumably be carried out for $2\alpha$ MZM-s for positive integer values of $\alpha$.

\section{Additional information regarding the DMRG simulations}
\label{extra_DMRG}
In this section, we provide additional information regarding the DMRG simulations and additional data which were not presented in the main-text for brevity. We simulated the JW transformed model describing spins coupled to rotors after the gauge transformation, given in Eqs. (\ref{ham2maj_2a}, \ref{ham2maj_2b}). To simulate the rotors, a truncation of 9 states per site ($n_i = -4, -3, \ldots, 3, 4$) was used. Together with the spin, thus, the local Hilbert space at each site was 18. We set $E_g=0$ while performing the simulations. The values of $E_g$ obtained to get the location of the boundaries of the Mott lobes were obtained by computing the cost of adding a particle/hole to the system (see below). We discuss the case when $E_J=\rho = 0$; other cases were done similarly. Note that in contrast to the ordinary Bose-Hubbard model~\cite{Kuhner2000}, here the densities can take negative values too. This is because here the particle number corresponds to the {\it extra} number of fermions, which can be negative. Physically, this correponds to creating a hole in the superconducting condensate present on each island. 

\subsection{The Mott-lobes and the KT transition}
\label{extra_DMRG_KT}
We describe the procedure for obtaining the lobes whose boundaries indicate the closing of the charge-$e$ excitation gap. Similar analysis is done for the charge-$2e$ gap. 
\begin{figure}
\includegraphics[width = 0.5\textwidth]{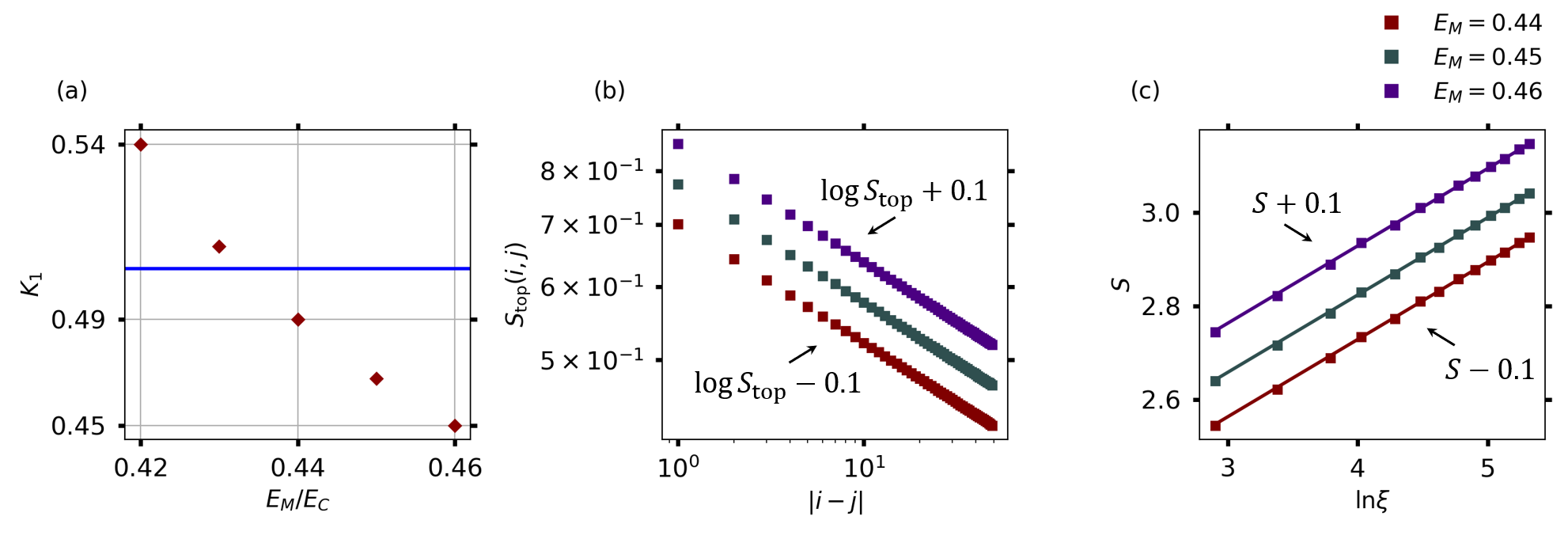}
\caption{\label{lobe_tip} iDMRG results for obtaining the location of the tip of the Mott lobe for $E_J = \rho = 0$. (a) Variation of the Luttinger parameter, $K_1$ (dark red diamonds), with the $E_M/E_C$. The phase-transition point is when $K_1$ crosses $1/2$ (blue horizontal line). The location of the transition is obtained by linear interpolation between the two points closest to $1/2$. (b) Variation of $S_{\rm{top}}(i,j)$ with $|i-j|$ in the charge-$e$ LL phase.  To accentuate the difference between the curves, we have added (subtracted) 0.1 to the y-axis values for the indigo (maroon) curves. The slope of the linear fit gives the Luttinger parameter $K_1$. (c) Variation of entanglement entropy ($S$) with the log of the correlation length ($\xi$). As expected for a critical system, $S$ scales linearly with $\ln\xi$, with the slope providing the central charge $c$. We have again added (subtracted) 0.1 to the y-axis values for the indigo (maroon) curves to make them distinct on this scale. }
\end{figure}

The ground state in the MI phase zero filling was obtained by starting with a state: 
\begin{equation}
|\psi_{\rm{init}}(0)\rangle = \bigotimes_{i=1}^L|0,\downarrow\rangle_i,
\end{equation}
where $|\downarrow\rangle_i$ corresponds to the spin-down state on the $i^{\rm{th}}$ island. Subsequently, the ground state was obtained by performing DMRG on this state while conserving the $U(1)$ charge $\sum_{i=1}^L\tilde{n}_i$ (see Sec. \ref{bosonic_model}). Denote the energy of the ground state so obtained by ${\cal E}_g(0)$, where the $0$ denotes that $\langle \sum_{i=1}^L \tilde{n}_i\rangle = 0$ for the corresponding state. The boundaries of the lobes for different fillings were obtained by looking at the cost of adding an particle or a hole to the system and setting it equal to the gate voltage. For the upper (lower) boundary of the lobe, first, a particle (hole) was created by applying the operator $\tilde{b}_j^\dagger (\tilde{b}_j)$ on the state $|\psi_{\rm{init}}(0)\rangle$, where $j\simeq L/2$. Subsequently, DMRG was used to get the energy starting from this state. Denoting the particle (hole) case by ${\cal E}_g(\pm1)$, we arrive at 
\begin{align}
 {\cal E}_g(+1)- {\cal E}_g(0) &= E_g^p(0),\nonumber\\
{\cal E}_g(0) - {\cal E}_g(-1) &= E_g^h(0),
\end{align}
where $E_g^{p(h)}$ denotes the value of the gate-voltage at the boundary upper (lower) boundary of the lobe. This way, the gap to the first-excited state of the Hamiltonian given in Eqs. (\ref{ham2maj_2a}, \ref{ham2maj_2b}) is zero as the system transitions from a gapped MI to a gapless charge-$e$ LL phase. While this way of obtaining the phase-boundary is sufficient for upper and lower sides of the Mott-lobes, the location of the tip of the lobe requires more work~\cite{Kuhner2000}. This is because the phase-transition at the sides and the tips of the lobes are of a fundamentally different nature, the latter being of KT type. The energy gap closes much more slowly (exponentially) with change of the parameter $E_M/E_C$ across the tip of the lobe. To ensure improved accuracy, we compute the change in the Luttinger parameter $K_1$. At the KT transition-point, $K_1=1/2$. The results for $E_J = \rho = 0$ are shown in Fig.~\ref{lobe_tip} for the phase-transition across the tip of the lobe. The maximum bond-dimension was chosen to be 280 to keep truncation errors of the order of $10^{-8}$. This was chosen to ensure that the correlation functions computed up to a distance of $|i-j| = 50$ were free of any truncation effects. For larger $|i-j|$, larger bond-dimensions will be necessary. The tip of the lobe is obtained by interpolating between the two points closest to 1/2 (see top left panel). This yields the critical point to be $E_M/E_C = 0.432\pm 0.002$. In the top right panel, the linear dependence of $\ln S_{\rm{top}}$ with $|i-j|$ is shown for three of the largest of values of $E_M/E_C$ of the top left panel. The central charge ($c$) is obtained by looking at the scaling of the entropy $S$ with $\ln\xi$, where $\xi$ is the correlation length of the system. In the scaling regime, for a gapless system, $S$ scales linearly with $\ln\xi$ with a slope of $c/6$.~\cite{Calabrese2009} The variation of $S$ with $\ln\xi$ for the three largest values of $E_M/E_C$ is shown in the bottom right panel. Similar results were obtained to infer the phase-transition for the $E_M = 0$ case with $S_b$ evaluated instead of $S_{\rm{top}}$; we do not show these for brevity. 

\subsection{The Ising transition}
\label{DMRG_Ising}
In this appendix, we provide additional numerical evidence capturing the Ising phase-transition separating the charge-$e$ LL phase from the other two phases. The complexity of the numerical simulations increases substantially to fully capture this phase-transition for the following reasons. First, the charge-$e$ excitation gap is much smaller than unity in the charge-$2e$ LL phase. As shown in Fig.~\ref{lobe_2}, this gap can be as small as $\sim10^{-3}-10^{-4}$. As a consequence, the system sizes needed to reliably obtain the aforementioned excitation gaps are rather large. To obtain the lobe of the right panel of Fig.~\ref{lobe_2}, system sizes up to 400 was simulated. Furthermore, for the parameters $E_J/E_C, E_M/E_C$, the ground state is highly entangled. As a result, to obtain truncation errors of the order of $10^{-8}$, a maximum bond-dimension of 400 was necessary. Second, we verified the central charge across the transition by using iDMRG and computing the scaling of entanglement entropy vs. the logarithm of the correlation length. To reach the regime where we could reliably capture the Ising transition, we needed to simulate up to a maximum bond-dimension of 1050 with truncation errors of the order of $\sim10^{-8}$. 

\bibliography{library_1}

\begin{thebibliography}{57}%
\makeatletter
\providecommand \@ifxundefined [1]{%
 \@ifx{#1\undefined}
}%
\providecommand \@ifnum [1]{%
 \ifnum #1\expandafter \@firstoftwo
 \else \expandafter \@secondoftwo
 \fi
}%
\providecommand \@ifx [1]{%
 \ifx #1\expandafter \@firstoftwo
 \else \expandafter \@secondoftwo
 \fi
}%
\providecommand \natexlab [1]{#1}%
\providecommand \enquote  [1]{``#1''}%
\providecommand \bibnamefont  [1]{#1}%
\providecommand \bibfnamefont [1]{#1}%
\providecommand \citenamefont [1]{#1}%
\providecommand \href@noop [0]{\@secondoftwo}%
\providecommand \href [0]{\begingroup \@sanitize@url \@href}%
\providecommand \@href[1]{\@@startlink{#1}\@@href}%
\providecommand \@@href[1]{\endgroup#1\@@endlink}%
\providecommand \@sanitize@url [0]{\catcode `\\12\catcode `\$12\catcode
  `\&12\catcode `\#12\catcode `\^12\catcode `\_12\catcode `\%12\relax}%
\providecommand \@@startlink[1]{}%
\providecommand \@@endlink[0]{}%
\providecommand \url  [0]{\begingroup\@sanitize@url \@url }%
\providecommand \@url [1]{\endgroup\@href {#1}{\urlprefix }}%
\providecommand \urlprefix  [0]{URL }%
\providecommand \Eprint [0]{\href }%
\providecommand \doibase [0]{http://dx.doi.org/}%
\providecommand \selectlanguage [0]{\@gobble}%
\providecommand \bibinfo  [0]{\@secondoftwo}%
\providecommand \bibfield  [0]{\@secondoftwo}%
\providecommand \translation [1]{[#1]}%
\providecommand \BibitemOpen [0]{}%
\providecommand \bibitemStop [0]{}%
\providecommand \bibitemNoStop [0]{.\EOS\space}%
\providecommand \EOS [0]{\spacefactor3000\relax}%
\providecommand \BibitemShut  [1]{\csname bibitem#1\endcsname}%
\let\auto@bib@innerbib\@empty
%</preamble>
\bibitem [{\citenamefont {Kitaev}(2003)}]{Kitaev2003}%
  \BibitemOpen
  \bibfield  {author} {\bibinfo {author} {\bibfnamefont {A.}~\bibnamefont
  {Kitaev}},\ }\href {\doibase 10.1016/S0003-4916(02)00018-0} {\bibfield
  {journal} {\bibinfo  {journal} {Ann. Phys. (NY)}\ }\textbf {\bibinfo {volume}
  {303}},\ \bibinfo {pages} {2} (\bibinfo {year} {2003})}\BibitemShut {NoStop}%
\bibitem [{\citenamefont {Kitaev}(2006)}]{Kitaev2006}%
  \BibitemOpen
  \bibfield  {author} {\bibinfo {author} {\bibfnamefont {A.}~\bibnamefont
  {Kitaev}},\ }\href {\doibase 10.1016/j.aop.2005.10.005} {\bibfield  {journal}
  {\bibinfo  {journal} {Ann. Phys. (NY)}\ }\textbf {\bibinfo {volume} {321}},\
  \bibinfo {pages} {2} (\bibinfo {year} {2006})}\BibitemShut {NoStop}%
\bibitem [{\citenamefont {Levin}\ and\ \citenamefont {Wen}(2005)}]{Levin2005}%
  \BibitemOpen
  \bibfield  {author} {\bibinfo {author} {\bibfnamefont {M.~A.}\ \bibnamefont
  {Levin}}\ and\ \bibinfo {author} {\bibfnamefont {X.-G.}\ \bibnamefont
  {Wen}},\ }\href {\doibase 10.1103/PhysRevB.71.045110} {\bibfield  {journal}
  {\bibinfo  {journal} {Phys. Rev. B}\ }\textbf {\bibinfo {volume} {71}},\
  \bibinfo {pages} {045110} (\bibinfo {year} {2005})}\BibitemShut {NoStop}%
\bibitem [{\citenamefont {Levin}\ and\ \citenamefont {Wen}(2006)}]{Levin2006}%
  \BibitemOpen
  \bibfield  {author} {\bibinfo {author} {\bibfnamefont {M.}~\bibnamefont
  {Levin}}\ and\ \bibinfo {author} {\bibfnamefont {X.-G.}\ \bibnamefont
  {Wen}},\ }\href {\doibase 10.1103/PhysRevLett.96.110405} {\bibfield
  {journal} {\bibinfo  {journal} {Phys. Rev. Lett.}\ }\textbf {\bibinfo
  {volume} {96}},\ \bibinfo {pages} {110405} (\bibinfo {year}
  {2006})}\BibitemShut {NoStop}%
\bibitem [{\citenamefont {Nayak}\ \emph {et~al.}(2008)\citenamefont {Nayak},
  \citenamefont {Simon}, \citenamefont {Stern}, \citenamefont {Freedman},\ and\
  \citenamefont {{Das Sarma}}}]{Nayak2008}%
  \BibitemOpen
  \bibfield  {author} {\bibinfo {author} {\bibfnamefont {C.}~\bibnamefont
  {Nayak}}, \bibinfo {author} {\bibfnamefont {S.~H.}\ \bibnamefont {Simon}},
  \bibinfo {author} {\bibfnamefont {A.}~\bibnamefont {Stern}}, \bibinfo
  {author} {\bibfnamefont {M.}~\bibnamefont {Freedman}}, \ and\ \bibinfo
  {author} {\bibfnamefont {S.}~\bibnamefont {{Das Sarma}}},\ }\href {\doibase
  10.1103/RevModPhys.80.1083} {\bibfield  {journal} {\bibinfo  {journal}
  {Reviews of Modern Physics}\ }\textbf {\bibinfo {volume} {80}},\ \bibinfo
  {pages} {1083} (\bibinfo {year} {2008})}\BibitemShut {NoStop}%
\bibitem [{\citenamefont {Read}\ and\ \citenamefont {Green}(2000)}]{Read2000}%
  \BibitemOpen
  \bibfield  {author} {\bibinfo {author} {\bibfnamefont {N.}~\bibnamefont
  {Read}}\ and\ \bibinfo {author} {\bibfnamefont {D.}~\bibnamefont {Green}},\
  }\href@noop {} {\bibfield  {journal} {\bibinfo  {journal} {Phys. Rev. B}\
  }\textbf {\bibinfo {volume} {61}},\ \bibinfo {pages} {10267} (\bibinfo {year}
  {2000})}\BibitemShut {NoStop}%
\bibitem [{\citenamefont {Bravyi}(2006)}]{Bravyi2006}%
  \BibitemOpen
  \bibfield  {author} {\bibinfo {author} {\bibfnamefont {S.}~\bibnamefont
  {Bravyi}},\ }\href {\doibase 10.1103/PhysRevA.73.042313} {\bibfield
  {journal} {\bibinfo  {journal} {Phys. Rev. A}\ }\textbf {\bibinfo {volume}
  {73}},\ \bibinfo {pages} {1} (\bibinfo {year} {2006})}\BibitemShut {NoStop}%
\bibitem [{\citenamefont {Kitaev}(2001)}]{Kitaev2001}%
  \BibitemOpen
  \bibfield  {author} {\bibinfo {author} {\bibfnamefont {A.}~\bibnamefont
  {Kitaev}},\ }\href@noop {} {\bibfield  {journal} {\bibinfo  {journal} {Phys.
  Usp. 44}\ }\textbf {\bibinfo {volume} {44}},\ \bibinfo {pages} {131}
  (\bibinfo {year} {2001})}\BibitemShut {NoStop}%
\bibitem [{\citenamefont {Xu}\ and\ \citenamefont {Fu}(2010)}]{Xu2010}%
  \BibitemOpen
  \bibfield  {author} {\bibinfo {author} {\bibfnamefont {C.}~\bibnamefont
  {Xu}}\ and\ \bibinfo {author} {\bibfnamefont {L.}~\bibnamefont {Fu}},\ }\href
  {\doibase 10.1103/PhysRevB.81.134435} {\bibfield  {journal} {\bibinfo
  {journal} {Phys. Rev. B}\ }\textbf {\bibinfo {volume} {81}},\ \bibinfo
  {pages} {1} (\bibinfo {year} {2010})}\BibitemShut {NoStop}%
\bibitem [{\citenamefont {Terhal}\ \emph {et~al.}(2012)\citenamefont {Terhal},
  \citenamefont {Hassler},\ and\ \citenamefont {Divincenzo}}]{Terhal2012}%
  \BibitemOpen
  \bibfield  {author} {\bibinfo {author} {\bibfnamefont {B.~M.}\ \bibnamefont
  {Terhal}}, \bibinfo {author} {\bibfnamefont {F.}~\bibnamefont {Hassler}}, \
  and\ \bibinfo {author} {\bibfnamefont {D.~P.}\ \bibnamefont {Divincenzo}},\
  }\href {\doibase 10.1103/PhysRevLett.108.260504} {\bibfield  {journal}
  {\bibinfo  {journal} {Phys. Rev. Lett.}\ }\textbf {\bibinfo {volume} {108}},\
  \bibinfo {pages} {1} (\bibinfo {year} {2012})}\BibitemShut {NoStop}%
\bibitem [{\citenamefont {Landau}\ \emph {et~al.}(2016)\citenamefont {Landau},
  \citenamefont {Plugge}, \citenamefont {Sela}, \citenamefont {Altland},
  \citenamefont {Albrecht},\ and\ \citenamefont {Egger}}]{Landau2016}%
  \BibitemOpen
  \bibfield  {author} {\bibinfo {author} {\bibfnamefont {L.~A.}\ \bibnamefont
  {Landau}}, \bibinfo {author} {\bibfnamefont {S.}~\bibnamefont {Plugge}},
  \bibinfo {author} {\bibfnamefont {E.}~\bibnamefont {Sela}}, \bibinfo {author}
  {\bibfnamefont {A.}~\bibnamefont {Altland}}, \bibinfo {author} {\bibfnamefont
  {S.~M.}\ \bibnamefont {Albrecht}}, \ and\ \bibinfo {author} {\bibfnamefont
  {R.}~\bibnamefont {Egger}},\ }\href {\doibase 10.1103/PhysRevLett.116.050501}
  {\bibfield  {journal} {\bibinfo  {journal} {Phys. Rev. Lett.}\ }\textbf
  {\bibinfo {volume} {116}},\ \bibinfo {pages} {1} (\bibinfo {year}
  {2016})}\BibitemShut {NoStop}%
\bibitem [{\citenamefont {Karzig}\ \emph {et~al.}(2017)\citenamefont {Karzig},
  \citenamefont {Knapp}, \citenamefont {Lutchyn}, \citenamefont {Bonderson},
  \citenamefont {Hastings}, \citenamefont {Nayak}, \citenamefont {Alicea},
  \citenamefont {Flensberg}, \citenamefont {Plugge}, \citenamefont {Oreg},
  \citenamefont {Marcus},\ and\ \citenamefont {Freedman}}]{Karzig2016}%
  \BibitemOpen
  \bibfield  {author} {\bibinfo {author} {\bibfnamefont {T.}~\bibnamefont
  {Karzig}}, \bibinfo {author} {\bibfnamefont {C.}~\bibnamefont {Knapp}},
  \bibinfo {author} {\bibfnamefont {R.~M.}\ \bibnamefont {Lutchyn}}, \bibinfo
  {author} {\bibfnamefont {P.}~\bibnamefont {Bonderson}}, \bibinfo {author}
  {\bibfnamefont {M.~B.}\ \bibnamefont {Hastings}}, \bibinfo {author}
  {\bibfnamefont {C.}~\bibnamefont {Nayak}}, \bibinfo {author} {\bibfnamefont
  {J.}~\bibnamefont {Alicea}}, \bibinfo {author} {\bibfnamefont
  {K.}~\bibnamefont {Flensberg}}, \bibinfo {author} {\bibfnamefont
  {S.}~\bibnamefont {Plugge}}, \bibinfo {author} {\bibfnamefont
  {Y.}~\bibnamefont {Oreg}}, \bibinfo {author} {\bibfnamefont {C.~M.}\
  \bibnamefont {Marcus}}, \ and\ \bibinfo {author} {\bibfnamefont {M.~H.}\
  \bibnamefont {Freedman}},\ }\href {\doibase 10.1103/PhysRevB.95.235305}
  {\bibfield  {journal} {\bibinfo  {journal} {Phys. Rev. B}\ }\textbf {\bibinfo
  {volume} {95}},\ \bibinfo {pages} {235305} (\bibinfo {year}
  {2017})}\BibitemShut {NoStop}%
\bibitem [{\citenamefont {Roy}\ \emph {et~al.}(2017)\citenamefont {Roy},
  \citenamefont {Terhal},\ and\ \citenamefont {Hassler}}]{Roy2017}%
  \BibitemOpen
  \bibfield  {author} {\bibinfo {author} {\bibfnamefont {A.}~\bibnamefont
  {Roy}}, \bibinfo {author} {\bibfnamefont {B.~M.}\ \bibnamefont {Terhal}}, \
  and\ \bibinfo {author} {\bibfnamefont {F.}~\bibnamefont {Hassler}},\ }\href
  {\doibase 10.1103/PhysRevLett.119.180508} {\bibfield  {journal} {\bibinfo
  {journal} {Phys. Rev. Lett.}\ }\textbf {\bibinfo {volume} {119}},\ \bibinfo
  {pages} {180508} (\bibinfo {year} {2017})}\BibitemShut {NoStop}%
\bibitem [{\citenamefont {Fidkowski}\ \emph {et~al.}(2011)\citenamefont
  {Fidkowski}, \citenamefont {Lutchyn}, \citenamefont {Nayak},\ and\
  \citenamefont {Fisher}}]{Fidkowski2011}%
  \BibitemOpen
  \bibfield  {author} {\bibinfo {author} {\bibfnamefont {L.}~\bibnamefont
  {Fidkowski}}, \bibinfo {author} {\bibfnamefont {R.~M.}\ \bibnamefont
  {Lutchyn}}, \bibinfo {author} {\bibfnamefont {C.}~\bibnamefont {Nayak}}, \
  and\ \bibinfo {author} {\bibfnamefont {M.~P.~A.}\ \bibnamefont {Fisher}},\
  }\href {\doibase 10.1103/PhysRevB.84.195436} {\bibfield  {journal} {\bibinfo
  {journal} {Phys. Rev. B}\ }\textbf {\bibinfo {volume} {84}},\ \bibinfo
  {pages} {195436} (\bibinfo {year} {2011})}\BibitemShut {NoStop}%
\bibitem [{\citenamefont {Sau}\ \emph {et~al.}(2011)\citenamefont {Sau},
  \citenamefont {Halperin}, \citenamefont {Flensberg},\ and\ \citenamefont
  {Das~Sarma}}]{Sau2011}%
  \BibitemOpen
  \bibfield  {author} {\bibinfo {author} {\bibfnamefont {J.~D.}\ \bibnamefont
  {Sau}}, \bibinfo {author} {\bibfnamefont {B.~I.}\ \bibnamefont {Halperin}},
  \bibinfo {author} {\bibfnamefont {K.}~\bibnamefont {Flensberg}}, \ and\
  \bibinfo {author} {\bibfnamefont {S.}~\bibnamefont {Das~Sarma}},\ }\href
  {\doibase 10.1103/PhysRevB.84.144509} {\bibfield  {journal} {\bibinfo
  {journal} {Phys. Rev. B}\ }\textbf {\bibinfo {volume} {84}},\ \bibinfo
  {pages} {144509} (\bibinfo {year} {2011})}\BibitemShut {NoStop}%
\bibitem [{\citenamefont {Cheng}\ and\ \citenamefont
  {Lutchyn}(2015)}]{Cheng2015}%
  \BibitemOpen
  \bibfield  {author} {\bibinfo {author} {\bibfnamefont {M.}~\bibnamefont
  {Cheng}}\ and\ \bibinfo {author} {\bibfnamefont {R.}~\bibnamefont
  {Lutchyn}},\ }\href {\doibase 10.1103/PhysRevB.92.134516} {\bibfield
  {journal} {\bibinfo  {journal} {Phys. Rev. B}\ }\textbf {\bibinfo {volume}
  {92}},\ \bibinfo {pages} {134516} (\bibinfo {year} {2015})}\BibitemShut
  {NoStop}%
\bibitem [{\citenamefont {Knapp}\ \emph {et~al.}(2018)\citenamefont {Knapp},
  \citenamefont {Karzig}, \citenamefont {Lutchyn},\ and\ \citenamefont
  {Nayak}}]{Knapp2018}%
  \BibitemOpen
  \bibfield  {author} {\bibinfo {author} {\bibfnamefont {C.}~\bibnamefont
  {Knapp}}, \bibinfo {author} {\bibfnamefont {T.}~\bibnamefont {Karzig}},
  \bibinfo {author} {\bibfnamefont {R.~M.}\ \bibnamefont {Lutchyn}}, \ and\
  \bibinfo {author} {\bibfnamefont {C.}~\bibnamefont {Nayak}},\ }\href
  {\doibase 10.1103/PhysRevB.97.125404} {\bibfield  {journal} {\bibinfo
  {journal} {Phys. Rev. B}\ }\textbf {\bibinfo {volume} {97}},\ \bibinfo
  {pages} {125404} (\bibinfo {year} {2018})}\BibitemShut {NoStop}%
\bibitem [{\citenamefont {{Knapp}}\ \emph {et~al.}(2019)\citenamefont
  {{Knapp}}, \citenamefont {{V{\"a}yrynen}},\ and\ \citenamefont
  {{Lutchyn}}}]{Knapp2019}%
  \BibitemOpen
  \bibfield  {author} {\bibinfo {author} {\bibfnamefont {C.}~\bibnamefont
  {{Knapp}}}, \bibinfo {author} {\bibfnamefont {J.~I.}\ \bibnamefont
  {{V{\"a}yrynen}}}, \ and\ \bibinfo {author} {\bibfnamefont {R.~M.}\
  \bibnamefont {{Lutchyn}}},\ }\href@noop {} {\bibfield  {journal} {\bibinfo
  {journal} {arXiv e-prints}\ ,\ \bibinfo {eid} {arXiv:1909.10521}} (\bibinfo
  {year} {2019})},\ \Eprint {http://arxiv.org/abs/1909.10521} {arXiv:1909.10521
  [cond-mat.mes-hall]} \BibitemShut {NoStop}%
\bibitem [{\citenamefont {Ortiz}\ \emph {et~al.}(2014)\citenamefont {Ortiz},
  \citenamefont {Dukelsky}, \citenamefont {Cobanera}, \citenamefont {Esebbag},\
  and\ \citenamefont {Beenakker}}]{Ortiz2014}%
  \BibitemOpen
  \bibfield  {author} {\bibinfo {author} {\bibfnamefont {G.}~\bibnamefont
  {Ortiz}}, \bibinfo {author} {\bibfnamefont {J.}~\bibnamefont {Dukelsky}},
  \bibinfo {author} {\bibfnamefont {E.}~\bibnamefont {Cobanera}}, \bibinfo
  {author} {\bibfnamefont {C.}~\bibnamefont {Esebbag}}, \ and\ \bibinfo
  {author} {\bibfnamefont {C.}~\bibnamefont {Beenakker}},\ }\href {\doibase
  10.1103/PhysRevLett.113.267002} {\bibfield  {journal} {\bibinfo  {journal}
  {Phys. Rev. Lett.}\ }\textbf {\bibinfo {volume} {113}},\ \bibinfo {pages}
  {267002} (\bibinfo {year} {2014})}\BibitemShut {NoStop}%
\bibitem [{\citenamefont {Iemini}\ \emph {et~al.}(2015)\citenamefont {Iemini},
  \citenamefont {Mazza}, \citenamefont {Rossini}, \citenamefont {Fazio},\ and\
  \citenamefont {Diehl}}]{Iemini2015}%
  \BibitemOpen
  \bibfield  {author} {\bibinfo {author} {\bibfnamefont {F.}~\bibnamefont
  {Iemini}}, \bibinfo {author} {\bibfnamefont {L.}~\bibnamefont {Mazza}},
  \bibinfo {author} {\bibfnamefont {D.}~\bibnamefont {Rossini}}, \bibinfo
  {author} {\bibfnamefont {R.}~\bibnamefont {Fazio}}, \ and\ \bibinfo {author}
  {\bibfnamefont {S.}~\bibnamefont {Diehl}},\ }\href {\doibase
  10.1103/PhysRevLett.115.156402} {\bibfield  {journal} {\bibinfo  {journal}
  {Phys. Rev. Lett.}\ }\textbf {\bibinfo {volume} {115}},\ \bibinfo {pages}
  {156402} (\bibinfo {year} {2015})}\BibitemShut {NoStop}%
\bibitem [{\citenamefont {Fisher}\ \emph {et~al.}(1989)\citenamefont {Fisher},
  \citenamefont {Weichman}, \citenamefont {Grinstein},\ and\ \citenamefont
  {Fisher}}]{Fisher1989}%
  \BibitemOpen
  \bibfield  {author} {\bibinfo {author} {\bibfnamefont {M.~P.~A.}\
  \bibnamefont {Fisher}}, \bibinfo {author} {\bibfnamefont {P.~B.}\
  \bibnamefont {Weichman}}, \bibinfo {author} {\bibfnamefont {G.}~\bibnamefont
  {Grinstein}}, \ and\ \bibinfo {author} {\bibfnamefont {D.~S.}\ \bibnamefont
  {Fisher}},\ }\href {\doibase 10.1103/PhysRevB.40.546} {\bibfield  {journal}
  {\bibinfo  {journal} {Phys. Rev. B}\ }\textbf {\bibinfo {volume} {40}},\
  \bibinfo {pages} {546} (\bibinfo {year} {1989})}\BibitemShut {NoStop}%
\bibitem [{\citenamefont {Fazio}\ and\ \citenamefont {van~der
  Zant}(2001)}]{Fazio2001}%
  \BibitemOpen
  \bibfield  {author} {\bibinfo {author} {\bibfnamefont {R.}~\bibnamefont
  {Fazio}}\ and\ \bibinfo {author} {\bibfnamefont {H.}~\bibnamefont {van~der
  Zant}},\ }\href {\doibase 10.1016/S0370-1573(01)00022-9} {\bibfield
  {journal} {\bibinfo  {journal} {Phys. Rep.}\ }\textbf {\bibinfo {volume}
  {355}},\ \bibinfo {pages} {235} (\bibinfo {year} {2001})}\BibitemShut
  {NoStop}%
\bibitem [{\citenamefont {Bradley}\ and\ \citenamefont
  {Doniach}(1984)}]{Bradley1984}%
  \BibitemOpen
  \bibfield  {author} {\bibinfo {author} {\bibfnamefont {R.~M.}\ \bibnamefont
  {Bradley}}\ and\ \bibinfo {author} {\bibfnamefont {S.}~\bibnamefont
  {Doniach}},\ }\href {\doibase 10.1103/PhysRevB.30.1138} {\bibfield  {journal}
  {\bibinfo  {journal} {Phys. Rev. B}\ }\textbf {\bibinfo {volume} {30}},\
  \bibinfo {pages} {1138} (\bibinfo {year} {1984})}\BibitemShut {NoStop}%
\bibitem [{\citenamefont {Korshunov}(1989)}]{Korshunov1989}%
  \BibitemOpen
  \bibfield  {author} {\bibinfo {author} {\bibfnamefont {S.~E.}\ \bibnamefont
  {Korshunov}},\ }\bibfield  {booktitle} {\emph {\bibinfo {booktitle} {Soviet
  Physics - JETP (English Translation)}},\ }\href
  {http://inis.iaea.org/search/search.aspx?orig_q=RN:22024752} {\ \textbf
  {\bibinfo {volume} {68}},\ \bibinfo {pages} {609} (\bibinfo {year}
  {1989})}\BibitemShut {NoStop}%
\bibitem [{\citenamefont {Glazman}\ and\ \citenamefont
  {Larkin}(1997)}]{Glazman1997}%
  \BibitemOpen
  \bibfield  {author} {\bibinfo {author} {\bibfnamefont {L.~I.}\ \bibnamefont
  {Glazman}}\ and\ \bibinfo {author} {\bibfnamefont {A.~I.}\ \bibnamefont
  {Larkin}},\ }\href {\doibase 10.1103/PhysRevLett.79.3736} {\bibfield
  {journal} {\bibinfo  {journal} {Phys. Rev. Lett.}\ }\textbf {\bibinfo
  {volume} {79}},\ \bibinfo {pages} {3736} (\bibinfo {year}
  {1997})}\BibitemShut {NoStop}%
\bibitem [{\citenamefont {Kogut}(1979)}]{Kogut1979}%
  \BibitemOpen
  \bibfield  {author} {\bibinfo {author} {\bibfnamefont {J.~B.}\ \bibnamefont
  {Kogut}},\ }\href {\doibase 10.1103/RevModPhys.51.659} {\bibfield  {journal}
  {\bibinfo  {journal} {Reviews of Modern Physics}\ }\textbf {\bibinfo {volume}
  {51}},\ \bibinfo {pages} {659} (\bibinfo {year} {1979})}\BibitemShut
  {NoStop}%
\bibitem [{\citenamefont {Pollmann}\ and\ \citenamefont
  {Turner}(2012)}]{Pollmann2012}%
  \BibitemOpen
  \bibfield  {author} {\bibinfo {author} {\bibfnamefont {F.}~\bibnamefont
  {Pollmann}}\ and\ \bibinfo {author} {\bibfnamefont {A.~M.}\ \bibnamefont
  {Turner}},\ }\href {\doibase 10.1103/PhysRevB.86.125441} {\bibfield
  {journal} {\bibinfo  {journal} {Phys. Rev. B}\ }\textbf {\bibinfo {volume}
  {86}},\ \bibinfo {pages} {125441} (\bibinfo {year} {2012})}\BibitemShut
  {NoStop}%
\bibitem [{\citenamefont {Bahri}\ and\ \citenamefont
  {Vishwanath}(2014)}]{Bahri2014}%
  \BibitemOpen
  \bibfield  {author} {\bibinfo {author} {\bibfnamefont {Y.}~\bibnamefont
  {Bahri}}\ and\ \bibinfo {author} {\bibfnamefont {A.}~\bibnamefont
  {Vishwanath}},\ }\href {\doibase 10.1103/PhysRevB.89.155135} {\bibfield
  {journal} {\bibinfo  {journal} {Phys. Rev. B}\ }\textbf {\bibinfo {volume}
  {89}},\ \bibinfo {pages} {155135} (\bibinfo {year} {2014})}\BibitemShut
  {NoStop}%
\bibitem [{\citenamefont {Roy}\ and\ \citenamefont {Hassler}(2018)}]{Roy2018a}%
  \BibitemOpen
  \bibfield  {author} {\bibinfo {author} {\bibfnamefont {A.}~\bibnamefont
  {Roy}}\ and\ \bibinfo {author} {\bibfnamefont {F.}~\bibnamefont {Hassler}},\
  }\href {\doibase 10.1103/PhysRevB.97.024512} {\bibfield  {journal} {\bibinfo
  {journal} {Phys. Rev. B}\ }\textbf {\bibinfo {volume} {97}},\ \bibinfo
  {pages} {024512} (\bibinfo {year} {2018})}\BibitemShut {NoStop}%
\bibitem [{\citenamefont {Lee}\ and\ \citenamefont
  {Grinstein}(1985)}]{Lee1985}%
  \BibitemOpen
  \bibfield  {author} {\bibinfo {author} {\bibfnamefont {D.~H.}\ \bibnamefont
  {Lee}}\ and\ \bibinfo {author} {\bibfnamefont {G.}~\bibnamefont
  {Grinstein}},\ }\href {\doibase 10.1103/PhysRevLett.55.541} {\bibfield
  {journal} {\bibinfo  {journal} {Phys. Rev. Lett.}\ }\textbf {\bibinfo
  {volume} {55}},\ \bibinfo {pages} {541} (\bibinfo {year} {1985})}\BibitemShut
  {NoStop}%
\bibitem [{\citenamefont {Serna}\ \emph {et~al.}(2017)\citenamefont {Serna},
  \citenamefont {Chalker},\ and\ \citenamefont {Fendley}}]{Serna2017}%
  \BibitemOpen
  \bibfield  {author} {\bibinfo {author} {\bibfnamefont {P.}~\bibnamefont
  {Serna}}, \bibinfo {author} {\bibfnamefont {J.~T.}\ \bibnamefont {Chalker}},
  \ and\ \bibinfo {author} {\bibfnamefont {P.}~\bibnamefont {Fendley}},\ }\href
  {\doibase 10.1088/1751-8121/aa89a1} {\bibfield  {journal} {\bibinfo
  {journal} {Journal of Physics A: Mathematical and Theoretical}\ }\textbf
  {\bibinfo {volume} {50}},\ \bibinfo {pages} {424003} (\bibinfo {year}
  {2017})}\BibitemShut {NoStop}%
\bibitem [{\citenamefont {Fradkin}\ and\ \citenamefont
  {Shenker}(1979)}]{Fradkin1979}%
  \BibitemOpen
  \bibfield  {author} {\bibinfo {author} {\bibfnamefont {E.}~\bibnamefont
  {Fradkin}}\ and\ \bibinfo {author} {\bibfnamefont {S.~H.}\ \bibnamefont
  {Shenker}},\ }\href {\doibase 10.1103/PhysRevD.19.3682} {\bibfield  {journal}
  {\bibinfo  {journal} {Phys. Rev. D}\ }\textbf {\bibinfo {volume} {19}},\
  \bibinfo {pages} {3682} (\bibinfo {year} {1979})}\BibitemShut {NoStop}%
\bibitem [{\citenamefont {Senthil}\ and\ \citenamefont
  {Fisher}(2000)}]{Senthil2000}%
  \BibitemOpen
  \bibfield  {author} {\bibinfo {author} {\bibfnamefont {T.}~\bibnamefont
  {Senthil}}\ and\ \bibinfo {author} {\bibfnamefont {M.~P.~A.}\ \bibnamefont
  {Fisher}},\ }\href {\doibase 10.1103/PhysRevB.62.7850} {\bibfield  {journal}
  {\bibinfo  {journal} {Phys. Rev. B}\ }\textbf {\bibinfo {volume} {62}},\
  \bibinfo {pages} {7850} (\bibinfo {year} {2000})}\BibitemShut {NoStop}%
\bibitem [{\citenamefont {Fu}(2010)}]{Fu2010}%
  \BibitemOpen
  \bibfield  {author} {\bibinfo {author} {\bibfnamefont {L.}~\bibnamefont
  {Fu}},\ }\href {\doibase 10.1103/PhysRevLett.104.056402} {\bibfield
  {journal} {\bibinfo  {journal} {Phys. Rev. Lett.}\ }\textbf {\bibinfo
  {volume} {104}},\ \bibinfo {pages} {1} (\bibinfo {year} {2010})}\BibitemShut
  {NoStop}%
\bibitem [{Note1()}]{Note1}%
  \BibitemOpen
  \bibinfo {note} {These operators are related to the Fredenhagen-Marcu
  operators of coventional lattice gauge theory~\cite {Fredenhagen1983,
  Gregor2011, Ziesen2019}}\BibitemShut {NoStop}%
\bibitem [{\citenamefont {van Heck}\ \emph {et~al.}(2012)\citenamefont {van
  Heck}, \citenamefont {Akhmerov}, \citenamefont {Hassler}, \citenamefont
  {Burrello},\ and\ \citenamefont {Beenakker}}]{VanHeck2012}%
  \BibitemOpen
  \bibfield  {author} {\bibinfo {author} {\bibfnamefont {B.}~\bibnamefont {van
  Heck}}, \bibinfo {author} {\bibfnamefont {A.~R.}\ \bibnamefont {Akhmerov}},
  \bibinfo {author} {\bibfnamefont {F.}~\bibnamefont {Hassler}}, \bibinfo
  {author} {\bibfnamefont {M.}~\bibnamefont {Burrello}}, \ and\ \bibinfo
  {author} {\bibfnamefont {C.~W.~J.}\ \bibnamefont {Beenakker}},\ }\href
  {\doibase 10.1088/1367-2630/14/3/035019} {\bibfield  {journal} {\bibinfo
  {journal} {New J. Phys.}\ }\textbf {\bibinfo {volume} {14}},\ \bibinfo
  {pages} {035019} (\bibinfo {year} {2012})}\BibitemShut {NoStop}%
\bibitem [{Note2()}]{Note2}%
  \BibitemOpen
  \bibinfo {note} {This transformation can be viewed as a different way to
  count charges on each island. Prior to the transformation, the spins carried
  no charge, but merely changed the fermion number parity. After this
  transformation, the charges are distributed across the operators $b_i,
  b_i^\dagger $ (create and destroy charge $2e$) and $\sigma _i^\pm $ (create
  and destroy charge $e$).}\BibitemShut {Stop}%
\bibitem [{\citenamefont {Sachdev}(2011)}]{Sachdev2011}%
  \BibitemOpen
  \bibfield  {author} {\bibinfo {author} {\bibfnamefont {S.}~\bibnamefont
  {Sachdev}},\ }\href {https://books.google.de/books?id=F3IkpxwpqSgC} {\emph
  {\bibinfo {title} {{Quantum Phase Transitions}}}}\ (\bibinfo  {publisher}
  {Cambridge University Press},\ \bibinfo {year} {2011})\BibitemShut {NoStop}%
\bibitem [{\citenamefont {Di~Francesco}\ \emph {et~al.}(1997)\citenamefont
  {Di~Francesco}, \citenamefont {Mathieu}, \citenamefont {S{\'e}n{\'e}chal},\
  and\ \citenamefont {Senechal}}]{diFrancesco1997}%
  \BibitemOpen
  \bibfield  {author} {\bibinfo {author} {\bibfnamefont {P.}~\bibnamefont
  {Di~Francesco}}, \bibinfo {author} {\bibfnamefont {P.}~\bibnamefont
  {Mathieu}}, \bibinfo {author} {\bibfnamefont {D.}~\bibnamefont
  {S{\'e}n{\'e}chal}}, \ and\ \bibinfo {author} {\bibfnamefont
  {D.}~\bibnamefont {Senechal}},\ }\href
  {https://books.google.de/books?id=keUrdME5rhIC} {\emph {\bibinfo {title}
  {Conformal Field Theory}}},\ Graduate Texts in Contemporary Physics\
  (\bibinfo  {publisher} {Springer},\ \bibinfo {year} {1997})\BibitemShut
  {NoStop}%
\bibitem [{\citenamefont {K\"uhner}\ \emph {et~al.}(2000)\citenamefont
  {K\"uhner}, \citenamefont {White},\ and\ \citenamefont
  {Monien}}]{Kuhner2000}%
  \BibitemOpen
  \bibfield  {author} {\bibinfo {author} {\bibfnamefont {T.~D.}\ \bibnamefont
  {K\"uhner}}, \bibinfo {author} {\bibfnamefont {S.~R.}\ \bibnamefont {White}},
  \ and\ \bibinfo {author} {\bibfnamefont {H.}~\bibnamefont {Monien}},\ }\href
  {\doibase 10.1103/PhysRevB.61.12474} {\bibfield  {journal} {\bibinfo
  {journal} {Phys. Rev. B}\ }\textbf {\bibinfo {volume} {61}},\ \bibinfo
  {pages} {12474} (\bibinfo {year} {2000})}\BibitemShut {NoStop}%
\bibitem [{\citenamefont {Chaikin}\ and\ \citenamefont
  {Lubensky}(2000)}]{Chaikin2000}%
  \BibitemOpen
  \bibfield  {author} {\bibinfo {author} {\bibfnamefont {P.}~\bibnamefont
  {Chaikin}}\ and\ \bibinfo {author} {\bibfnamefont {T.}~\bibnamefont
  {Lubensky}},\ }\href {https://books.google.de/books?id=P9YjNjzr9OIC} {\emph
  {\bibinfo {title} {Principles of Condensed Matter Physics}}}\ (\bibinfo
  {publisher} {Cambridge University Press},\ \bibinfo {year}
  {2000})\BibitemShut {NoStop}%
\bibitem [{\citenamefont {Hauschild}\ and\ \citenamefont
  {Pollmann}(2018)}]{Hauschild2018}%
  \BibitemOpen
  \bibfield  {author} {\bibinfo {author} {\bibfnamefont {J.}~\bibnamefont
  {Hauschild}}\ and\ \bibinfo {author} {\bibfnamefont {F.}~\bibnamefont
  {Pollmann}},\ }\href {\doibase 10.21468/SciPostPhysLectNotes.5} {\bibfield
  {journal} {\bibinfo  {journal} {SciPost Phys. Lect. Notes}\ ,\ \bibinfo
  {pages} {5}} (\bibinfo {year} {2018})}\BibitemShut {NoStop}%
\bibitem [{\citenamefont {Gu}(2010)}]{Gu2010}%
  \BibitemOpen
  \bibfield  {author} {\bibinfo {author} {\bibfnamefont {S.-J.}\ \bibnamefont
  {Gu}},\ }\href {\doibase 10.1142/S0217979210056335} {\bibfield  {journal}
  {\bibinfo  {journal} {International Journal of Modern Physics B}\ }\textbf
  {\bibinfo {volume} {24}},\ \bibinfo {pages} {4371} (\bibinfo {year}
  {2010})},\ \Eprint
  {http://arxiv.org/abs/https://doi.org/10.1142/S0217979210056335}
  {https://doi.org/10.1142/S0217979210056335} \BibitemShut {NoStop}%
\bibitem [{\citenamefont {Giamarchi}(2003)}]{Giamarchi2003}%
  \BibitemOpen
  \bibfield  {author} {\bibinfo {author} {\bibfnamefont {T.}~\bibnamefont
  {Giamarchi}},\ }\href {https://books.google.de/books?id=GVeuKZLGMZ0C} {\emph
  {\bibinfo {title} {Quantum Physics in One Dimension}}},\ International Series
  of Monographs on Physics\ (\bibinfo  {publisher} {Clarendon Press},\ \bibinfo
  {year} {2003})\BibitemShut {NoStop}%
\bibitem [{\citenamefont {Schmid}(1983)}]{Schmid1983}%
  \BibitemOpen
  \bibfield  {author} {\bibinfo {author} {\bibfnamefont {A.}~\bibnamefont
  {Schmid}},\ }\href {\doibase 10.1103/PhysRevLett.51.1506} {\bibfield
  {journal} {\bibinfo  {journal} {Phys. Rev. Lett.}\ }\textbf {\bibinfo
  {volume} {51}},\ \bibinfo {pages} {1506} (\bibinfo {year}
  {1983})}\BibitemShut {NoStop}%
\bibitem [{\citenamefont {Caldeira}\ and\ \citenamefont
  {Leggett}(1983)}]{Caldeira_Leggett_1983}%
  \BibitemOpen
  \bibfield  {author} {\bibinfo {author} {\bibfnamefont {A.~O.}\ \bibnamefont
  {Caldeira}}\ and\ \bibinfo {author} {\bibfnamefont {A.~J.}\ \bibnamefont
  {Leggett}},\ }\href {\doibase http://dx.doi.org/10.1016/0003-4916(83)90202-6}
  {\bibfield  {journal} {\bibinfo  {journal} {Ann. Phys. (NY)}\ }\textbf
  {\bibinfo {volume} {149}},\ \bibinfo {pages} {374} (\bibinfo {year}
  {1983})}\BibitemShut {NoStop}%
\bibitem [{\citenamefont {Kane}\ and\ \citenamefont
  {Fisher}(1992{\natexlab{a}})}]{Kane1992}%
  \BibitemOpen
  \bibfield  {author} {\bibinfo {author} {\bibfnamefont {C.~L.}\ \bibnamefont
  {Kane}}\ and\ \bibinfo {author} {\bibfnamefont {M.~P.~A.}\ \bibnamefont
  {Fisher}},\ }\href {\doibase 10.1103/PhysRevLett.68.1220} {\bibfield
  {journal} {\bibinfo  {journal} {Phys. Rev. Lett.}\ }\textbf {\bibinfo
  {volume} {68}},\ \bibinfo {pages} {1220} (\bibinfo {year}
  {1992}{\natexlab{a}})}\BibitemShut {NoStop}%
\bibitem [{\citenamefont {Kane}\ and\ \citenamefont
  {Fisher}(1992{\natexlab{b}})}]{Kane1992a}%
  \BibitemOpen
  \bibfield  {author} {\bibinfo {author} {\bibfnamefont {C.~L.}\ \bibnamefont
  {Kane}}\ and\ \bibinfo {author} {\bibfnamefont {M.~P.~A.}\ \bibnamefont
  {Fisher}},\ }\href {\doibase 10.1103/PhysRevB.46.15233} {\bibfield  {journal}
  {\bibinfo  {journal} {Phys. Rev. B}\ }\textbf {\bibinfo {volume} {46}},\
  \bibinfo {pages} {15233} (\bibinfo {year} {1992}{\natexlab{b}})}\BibitemShut
  {NoStop}%
\bibitem [{\citenamefont {Bonderson}\ and\ \citenamefont
  {Nayak}(2013)}]{Bonderson2013}%
  \BibitemOpen
  \bibfield  {author} {\bibinfo {author} {\bibfnamefont {P.}~\bibnamefont
  {Bonderson}}\ and\ \bibinfo {author} {\bibfnamefont {C.}~\bibnamefont
  {Nayak}},\ }\href {\doibase 10.1103/PhysRevB.87.195451} {\bibfield  {journal}
  {\bibinfo  {journal} {Phys. Rev. B}\ }\textbf {\bibinfo {volume} {87}},\
  \bibinfo {pages} {195451} (\bibinfo {year} {2013})}\BibitemShut {NoStop}%
\bibitem [{\citenamefont {Turner}\ \emph {et~al.}(2011)\citenamefont {Turner},
  \citenamefont {Pollmann},\ and\ \citenamefont {Berg}}]{Pollmann2011}%
  \BibitemOpen
  \bibfield  {author} {\bibinfo {author} {\bibfnamefont {A.~M.}\ \bibnamefont
  {Turner}}, \bibinfo {author} {\bibfnamefont {F.}~\bibnamefont {Pollmann}}, \
  and\ \bibinfo {author} {\bibfnamefont {E.}~\bibnamefont {Berg}},\ }\href
  {\doibase 10.1103/PhysRevB.83.075102} {\bibfield  {journal} {\bibinfo
  {journal} {Phys. Rev. B}\ }\textbf {\bibinfo {volume} {83}},\ \bibinfo
  {pages} {075102} (\bibinfo {year} {2011})}\BibitemShut {NoStop}%
\bibitem [{\citenamefont {Fidkowski}\ and\ \citenamefont
  {Kitaev}(2011)}]{Fidkowski2011a}%
  \BibitemOpen
  \bibfield  {author} {\bibinfo {author} {\bibfnamefont {L.}~\bibnamefont
  {Fidkowski}}\ and\ \bibinfo {author} {\bibfnamefont {A.}~\bibnamefont
  {Kitaev}},\ }\href {\doibase 10.1103/PhysRevB.83.075103} {\bibfield
  {journal} {\bibinfo  {journal} {Phys. Rev. B}\ }\textbf {\bibinfo {volume}
  {83}},\ \bibinfo {pages} {075103} (\bibinfo {year} {2011})}\BibitemShut
  {NoStop}%
\bibitem [{\citenamefont {Verresen}\ \emph {et~al.}(2017)\citenamefont
  {Verresen}, \citenamefont {Moessner},\ and\ \citenamefont
  {Pollmann}}]{Verresen2017}%
  \BibitemOpen
  \bibfield  {author} {\bibinfo {author} {\bibfnamefont {R.}~\bibnamefont
  {Verresen}}, \bibinfo {author} {\bibfnamefont {R.}~\bibnamefont {Moessner}},
  \ and\ \bibinfo {author} {\bibfnamefont {F.}~\bibnamefont {Pollmann}},\
  }\href {\doibase 10.1103/PhysRevB.96.165124} {\bibfield  {journal} {\bibinfo
  {journal} {Phys. Rev. B}\ }\textbf {\bibinfo {volume} {96}},\ \bibinfo
  {pages} {165124} (\bibinfo {year} {2017})}\BibitemShut {NoStop}%
\bibitem [{Note3()}]{Note3}%
  \BibitemOpen
  \bibinfo {note} {A different choice of the Jordan-Wigner string entails
  traversing all four MZM-s on each island in the order $a,b,d,c$ and the
  islands are traversed from left to right. This leads to the cluster
  model~\cite {Verresen2017}. However, not all the symmetries are easily
  apparent with this choice.}\BibitemShut {Stop}%
\bibitem [{\citenamefont {Calabrese}\ and\ \citenamefont
  {Cardy}(2009)}]{Calabrese2009}%
  \BibitemOpen
  \bibfield  {author} {\bibinfo {author} {\bibfnamefont {P.}~\bibnamefont
  {Calabrese}}\ and\ \bibinfo {author} {\bibfnamefont {J.}~\bibnamefont
  {Cardy}},\ }\href {\doibase 10.1088/1751-8113/42/50/504005} {\bibfield
  {journal} {\bibinfo  {journal} {J. Phys.}\ }\textbf {\bibinfo {volume}
  {A42}},\ \bibinfo {pages} {504005} (\bibinfo {year} {2009})},\ \Eprint
  {http://arxiv.org/abs/0905.4013} {arXiv:0905.4013 [cond-mat.stat-mech]}
  \BibitemShut {NoStop}%
%%CITATION = ARXIV:0905.4013;%%
\bibitem [{\citenamefont {Fredenhagen}\ and\ \citenamefont
  {Marcu}(1983)}]{Fredenhagen1983}%
  \BibitemOpen
  \bibfield  {author} {\bibinfo {author} {\bibfnamefont {K.}~\bibnamefont
  {Fredenhagen}}\ and\ \bibinfo {author} {\bibfnamefont {M.}~\bibnamefont
  {Marcu}},\ }\href {\doibase 10.1007/BF01206315} {\bibfield  {journal}
  {\bibinfo  {journal} {Communications in Mathematical Physics}\ }\textbf
  {\bibinfo {volume} {92}},\ \bibinfo {pages} {81} (\bibinfo {year}
  {1983})}\BibitemShut {NoStop}%
\bibitem [{\citenamefont {Gregor}\ \emph {et~al.}(2011)\citenamefont {Gregor},
  \citenamefont {Huse}, \citenamefont {Moessner},\ and\ \citenamefont
  {Sondhi}}]{Gregor2011}%
  \BibitemOpen
  \bibfield  {author} {\bibinfo {author} {\bibfnamefont {K.}~\bibnamefont
  {Gregor}}, \bibinfo {author} {\bibfnamefont {D.~A.}\ \bibnamefont {Huse}},
  \bibinfo {author} {\bibfnamefont {R.}~\bibnamefont {Moessner}}, \ and\
  \bibinfo {author} {\bibfnamefont {S.~L.}\ \bibnamefont {Sondhi}},\ }\href
  {\doibase 10.1088/1367-2630/13/2/025009} {\bibfield  {journal} {\bibinfo
  {journal} {New Journal of Physics}\ }\textbf {\bibinfo {volume} {13}},\
  \bibinfo {pages} {025009} (\bibinfo {year} {2011})}\BibitemShut {NoStop}%
\bibitem [{\citenamefont {Ziesen}\ \emph {et~al.}(2019)\citenamefont {Ziesen},
  \citenamefont {Hassler},\ and\ \citenamefont {Roy}}]{Ziesen2019}%
  \BibitemOpen
  \bibfield  {author} {\bibinfo {author} {\bibfnamefont {A.}~\bibnamefont
  {Ziesen}}, \bibinfo {author} {\bibfnamefont {F.}~\bibnamefont {Hassler}}, \
  and\ \bibinfo {author} {\bibfnamefont {A.}~\bibnamefont {Roy}},\ }\href
  {\doibase 10.1103/PhysRevB.100.104508} {\bibfield  {journal} {\bibinfo
  {journal} {Phys. Rev. B}\ }\textbf {\bibinfo {volume} {100}},\ \bibinfo
  {pages} {104508} (\bibinfo {year} {2019})}\BibitemShut {NoStop}%
\end{thebibliography}%

\end{document}